\documentclass{article}
\usepackage{arxiv}

\usepackage{amsmath,amsfonts,amssymb}
\usepackage{mathtools}
\usepackage[hidelinks]{hyperref}
\usepackage{algorithmic}
\usepackage{algorithm}
\usepackage{array}
\usepackage{adjustbox}
\usepackage[caption=false,font=normalsize,labelfont=sf,textfont=sf]{subfig}
\usepackage{relsize}
\usepackage{textcomp}
\usepackage{stfloats}
\usepackage{url}
\usepackage{verbatim}
\usepackage{graphicx}
\usepackage{natbib}
\usepackage{doi}
\usepackage{xcolor}
\usepackage{paralist}
\usepackage{blindtext}
\usepackage{multicol}
\usepackage{multirow}
\usepackage[subnum]{cases}

\newcount\Comments
\usepackage{color}
\newcommand{\kibitz}[2]{\ifnum\Comments=0\textcolor{#1}{#2}\fi}

\title{Safety, Mobility, and Environmental Impacts of Driver-Assistance-Enabled Electric Vehicles: An Empirical Study}

\author{
Gabriel Geffen\\
	Civil and Architectural Engineering and Mechanics\\
	The University of Arizona\\
	\texttt{gabriel9@arizona.edu} \\
	\And
Jun Zhao\\
	Civil and Architectural Engineering and Mechanics\\
	The University of Arizona\\
	\texttt{zhaojun@arizona.edu} \\
	\And
	Mingfeng Shang\\
	Civil and Architectural Engineering and Mechanics\\
	The University of Arizona\\
	\texttt{mfshang@arizona.edu} \\
	\And
	Shian Wang\\
	Civil, Environmental, and Architectural Engineering\\
	The University of Kansas\\
	\texttt{shian.wang@ku.edu}\\
	\And
  Yao-Jan Wu\\
	Civil and Architectural Engineering and Mechanics\\
	The University of Arizona\\
	\texttt{yaojan@arizona.edu} \\
}

\renewcommand{\shorttitle}{Safety, Mobility, and Environmental Impacts of Driver-Assistance-Enabled Electric Vehicles: An Empirical Study}

\begin{document}\sloppy
\maketitle

\begin{abstract}
The advancement of vehicle automation and the growing adoption of electric vehicles (EVs) are reshaping transportation systems. While fully automated vehicles are expected to improve traffic stability, efficiency, and sustainability, recent studies suggest that partially automated vehicles, such as those equipped with adaptive cruise control (ACC), may adversely affect traffic flow. These drawbacks may not extend to ACC-enabled EVs due to their distinct mechanical characteristics, including regenerative braking and smoother torque delivery. As a result, the impacts of EVs operating under ACC remain insufficiently understood.

To address this gap, this study develops an empirical framework using the OpenACC dataset to compare ACC-enabled EVs and internal combustion engine vehicles. Dynamic time warping aligns comparable lead-vehicle trajectories. Results show that EVs exhibit smoother speed profiles, lower speed variability, and shorter spacing, leading to higher efficiency. EVs reduce critical safety events by over 85\% and lower platoon-level emissions by up to 26.2\%.
\end{abstract}


\section{Introduction}\label{sec:intro}

The emergence, evolution, and advancement of vehicle automation are widely adopted as a key contributor to future transportation improvements~\citep{wang2022policy}. The Society of Automotive Engineers (SAE) defines five levels of driving automation, ranging from Level~1, which includes driver assistance features such as adaptive cruise control (ACC), to Level~5, which denotes full autonomy without human intervention. Commercially available vehicles equipped with ACC currently operate at Level~1 or Level~2, where active driver supervision and attention are still required. 

Fully automated vehicles (AVs), corresponding to Level~5 automation, are expected to improve future traffic flow by enhancing stability, increasing throughput and parking efficiency, and reducing fuel consumption and emissions~\citep{qin2018stability,talebpour2016influence,wang2021park,davis2004effect,sun2022energy,aguilar2024energy}. While these benefits have been widely projected for fully automated vehicles, recent studies suggest that they may not hold for partially automated vehicles, such as those equipped with commercially available ACC systems. In particular, ACC-equipped vehicles have been found to potentially degrade traffic flow by reducing throughput~\citep{shang2021impacts,shang2023extending,shang2023capacity}, decreasing stability~\citep{gunter2020commercially}, and increasing fuel consumption and emissions~\citep{shang2023impacts}. 

In parallel, electric vehicles (EVs) have emerged as a fast-growing alternative to internal combustion engine vehicles (ICEVs), driven by environmental regulation, technological advancement, and consumer demand. Given that many commercially available EVs are equipped with ACC systems~\citep{makridis2021openacc}, it is essential to assess how electrification interacts with automation in shaping driving behavior. Although EVs and ICEVs share fundamental operational principles, their powertrains differ significantly in performance characteristics~\citep{he2023real,acedo2025impacts,he2025connectivity}. Specifically, EVs offer regenerative braking, which enables smoother and more immediate deceleration across a wider range of speeds, and electric motors deliver maximum torque instantly from a standstill, allowing EVs to respond more quickly to traffic conditions. Consequently, the mechanical distinctions between EVs and ICEVs may lead to fundamentally different behaviors in ACC-enabled driving.

While previous studies have examined differences in driving behavior between ACC-enabled EVs and ICEVs~\citep{lapardhaja2023unlocking}, the focus has primarily been on microscopic-level analyses, characterizing behavior in terms of acceleration, speed, time headway, and similar metrics. However, how these microscopic behaviors translate into macroscopic traffic dynamics, such as traffic stability, throughput, emissions, and fuel consumption, remains underexplored. Recent studies have made preliminary contributions to understanding the differences between EVs and ICEVs in terms of their impacts on macroscopic traffic flow. For example, Zare et al.~\citep{zare2024electric} calibrated car-following models for EVs using real-world traffic data, which were subsequently used by Aguilar et al.~\citep{aguilar2024impacts} to investigate the effects of EVs on traffic stability and energy consumption across varying penetration rates. While simulation-based studies are widely employed in macroscopic traffic analysis~\citep{shang2021impacts,talebpour2016influence}, they may not be able to capture the full complexity of real-world traffic conditions, including disturbances and noise inherent in actual driving scenarios, compared to empirical analyses~\citep{geffen2025evaluation}. With this in mind, this study proposes an empirical framework to examine the driving behavior of EVs and ICEVs in ACC-enabled scenarios. The use of real-world data enhances the accuracy of impact assessment and provides reliable results for understanding the effects of EVs and ICEVs on macroscopic traffic dynamics.

However, unlike simulation-based studies, empirical studies face unique challenges that simulations may overlook. A key requirement is access to high-fidelity datasets that include both ACC-enabled EVs and ICEVs. While commonly used datasets, such as MicroSimACC~\citep{yang2024microsimacc}, Massachusetts~\citep{li2021car}, and Vanderbilt~\citep{gunter2020commercially} experiments, have contributed valuable ACC-related data, they may not be well-suited for studying the differences between EVs and ICEVs due to limitations in experimental consistency and scope of design. Additionally, for a comparative study on car-following behavior between an ICEV and an EV, controlling the behavior of the lead vehicle is challenging and crucial. Without proper control of the lead vehicle with similar trajectories, significant comparison errors may arise, making it difficult to accurately capture the nuanced differences between ACC-enabled EVs and ICEVs.

To this end, the present study utilizes the OpenACC dataset~\citep{makridis2021openacc}, which offers a comprehensive and standardized platform for evaluating ACC performance across both EVs and ICEVs. The dataset includes scenarios where vehicles operate under consistent experimental conditions, with data collected from a diverse range of commercially available ACC vehicles. The OpenACC dataset also offers detailed information essential for conducting empirical analyses of macroscopic traffic flow and is widely used in studies exploring autonomous driving behavior~\citep{chen2024safety,huang2023characterizing}. To address discrepancies in lead vehicle trajectories, dynamic time warping (DTW) has been utilized to identify and align comparable car-following scenarios~\citep{sousa2020vehicle,hu2023spatio}. DTW is a time-series alignment technique that measures similarity between two sequences that may vary in speed or length, enabling the matching of similar lead vehicle trajectories by aligning patterns in time. While not specific to EVs or ICEVs, DTW provides a promising and robust mechanism for controlling lead vehicle behavior and facilitates rigorous comparative analysis in empirical studies.

Building on these methodological advances, the contributions of this study are threefold:
\begin{itemize}
    \item Unlike prior research that largely relies on simulation-based approaches, we provide an empirical investigation using the high-fidelity OpenACC dataset, which allows us to directly capture real-world differences in car-following behavior between ACC-enabled EVs and ICEVs.
    \item We develop a structured framework for assessing the behavior of ACC-enabled EVs and ICEVs in terms of traffic safety, efficiency, and sustainability. Importantly, this framework is not limited to the present study and can be applied to a wide range of empirical and simulation-based investigations of vehicle behavior within the transportation community.
    \item We employ a trajectory similarity method based on DTW to identify cases of comparable lead vehicle behavior. The application of DTW ensures rigorous and equivalent comparisons of car-following behavior across EVs and ICEVs in empirical studies.
\end{itemize}

The remainder of this article is organized as follows. Initially, we provide a comprehensive literature review on commercially available ACC datasets, empirical studies on car-following behavior, and prior approaches to processing such datasets in Section~\ref{sec:related-work}. Next, we introduce the experimental dataset and describe the preprocessing procedures in Section~\ref{sec:dataset}. We then present the evaluation framework used to assess impacts on safety, mobility, and emissions in Section~\ref{sec:framework}. This is followed by the presentation of key findings and results in Section~\ref{sec:results}. Finally, we conclude the article and outline directions for future research in Section~\ref{sec:conclusions}.

\section{Related Work}\label{sec:related-work}

In this section, we conduct a comprehensive review of the three key areas relevant to this study. Specifically, we begin by reviewing publicly available datasets commonly used to capture car-following behavior. Next, we examine prior literature on empirical studies on macroscopic traffic analysis and microscopic car-following studies. Finally, we review diverse trajectory comparison methods, such as DTW, and their applications in transportation research.

\subsection{Car-Following Datasets}\label{sec:car-following-datasets}

In recent decades, several experimental datasets have been made publicly available to capture the driving behavior of ACC-equipped vehicles. For example, the Partners for Advanced Transportation Technology (PATH) dataset, developed by the University of California, Berkeley~\citep{milanes2014modeling}, is one of the earliest and most widely used datasets for investigating ACC behavior. However, the PATH dataset is limited, as it focuses on a small number of implementations collected from individual vehicles. Similarly, the Federal Highway Administration (FHWA) and the University of South Florida (USF) have collected datasets using one or two vehicles across multiple trials, typically under fixed-speed or simplified leader-following scenarios~\citep{james2019characterizing,shi2021empirical}. Given these datasets involve a small number of vehicles and controlled experimental conditions, they may be limited in their ability to represent real-world traffic dynamics, including multi-vehicle interactions and disturbance propagation in larger platoons.

Recently, the Massachusetts datasets~\citep{li2021car} and Vanderbilt datasets~\citep{gunter2020commercially} have introduced more diverse data collections, incorporating different vehicle brands, engine types, and configurations. Specifically, the Vanderbilt datasets include seven individual two-vehicle experiments conducted under various testing scenarios. While these experiments provide valuable insights into the driving behavior of ACC-equipped vehicles, they primarily involve conventional ICEVs and do not include EVs, limiting their applicability to heterogeneous traffic conditions. Additionally, the testing conditions are often constrained to narrow speed ranges and steady-state car-following scenarios, with limited variation in vehicle types or ACC control systems.

With this in mind, the OpenACC dataset~\citep{makridis2021openacc} addresses these limitations, including a large number of vehicles and ACC systems across a diverse range of platoon sizes, driving conditions, and vehicle types, including both EVs and ICEVs. The inclusion of large platoons enables the analysis of multi-vehicle interactions and platoon-wide analyses, while the range of driving conditions allows for evaluating ACC performance under both urban and highway scenarios. Although the dataset was collected in Europe, its controlled, experiment-based nature makes it suitable for comparative analysis relevant to U.S. traffic contexts. By capturing vehicle behavior in both urban and highway environments, the OpenACC dataset fills critical gaps in the existing dataset landscape and enables insights that are more representative of real-world ACC performance across different vehicle technologies and traffic conditions. Therefore, this study leverages the OpenACC dataset to conduct a comprehensive empirical comparison between ACC-enabled EVs and ICEVs.

\subsection{Empirical Studies on the Impacts of Car-Following Behavior}\label{sec:car-following-studies}

Unlike simulation studies, empirical analysis enables direct observation of how vehicles and drivers respond to real-world traffic stimuli, offering greater realism and applicability for traffic flow analysis, safety evaluation, and environmental impact assessment~\citep{ali2020impact,talebpour2016influence}. Accordingly, many researchers have relied on high-resolution trajectory data and controlled field experiments to examine car-following behavior and to validate longitudinal control models~\citep{coifman2017critical}. Within this body of work, a growing subset of empirical studies has focused specifically on ACC and partially automated vehicles, recognizing that automation fundamentally alters longitudinal driving behavior relative to human-driven vehicles. Field experiments have shown that observed ACC response times are comparable to those of human drivers, yet often exceed the controller's intended time gap~\citep{li2021car}. Additionally, empirical analyses of vehicles with higher levels of automation (i.e., SAE Levels 4--5) indicate potential safety benefits, as these vehicles tend to maintain larger time headways and jam spacings than human-driven vehicles~\citep{hu2023autonomous}.

Empirical studies of ACC systems further demonstrate that automation introduces distinct trade-offs among safety, mobility, and traffic flow stability~\citep{shi2021empirical}. Using field experiments and high-resolution trajectory data, prior work has shown that commercial ACC implementations can alter longitudinal driving dynamics relative to human-driven vehicles, including changes in response time, headway selection, and speed variability~\citep{shi2021empirical,shi2022empirical}. Specifically, while some empirical analyses report improved disturbance attenuation under longer ACC headway settings, others observe amplification of speed oscillations as traffic perturbations propagate through ACC-equipped platoons, indicating that real-world ACC behavior may deviate from idealized controller designs.

Beyond ACC driving behavior, empirical evaluations have also examined the energy and emissions impacts of ACC operation~\citep{shi2022empirical}. Field studies shown in~\citep{shi2022empirical} suggest that, under certain operating conditions, ACC can reduce speed variability relative to human driving, which may lead to improvements in fuel efficiency. However, these benefits are highly sensitive to controller tuning parameters, traffic context, and vehicle dynamics~\citep{shi2022empirical}. Notably, most empirical analyses do not distinguish between vehicle propulsion types, and instead evaluate ACC-equipped vehicles without explicitly considering how powertrain differences may influence observed outcomes.

As a result, existing studies provide limited insight into how inherent powertrain characteristics shape longitudinal driving behavior under ACC. In particular, current empirical studies lack systematic comparisons that isolate behavioral differences arising from EV-specific characteristics, such as instant torque delivery or regenerative braking, relative to ICEVs operating under ACC. Quantifying these powertrain-related distinctions empirically is essential for accurately modeling and predicting the impacts of increasingly electrified and automated vehicle fleets on traffic flow dynamics and energy consumption.

\subsection{Trajectory Similarity Analysis}\label{sec:trajectory-similaity-litreview}

Empirical data analysis presents challenges in identifying similarity in lead vehicle behavior, prompting the use of various trajectory similarity measures in prior studies. Common metrics include Euclidean distance, Fréchet distance, Hausdorff distance, and DTW~\citep{sousa2020vehicle,hu2023spatio,tao2021comparative}, each suited to different data characteristics. While Euclidean distance is straightforward, it assumes perfect temporal alignment and thus struggles with time-shifted sequences. Fréchet and Hausdorff distances are more applicable to comparing spatial paths but are sensitive to sequence length mismatches and lack flexibility for temporal data. DTW, in contrast, non-linearly aligns sequences in time and handles variable-length inputs, making it well-suited for comparing vehicle trajectories when timing differs~\citep{sousa2020vehicle,hu2023spatio,tao2021comparative}.

Given its advantages, DTW has seen widespread use in transportation research, including driver behavior analysis, traffic monitoring, and trajectory-based prediction and anomaly detection. Unlike the studies that often focus on identifying abnormal patterns, the objective of this study is to compare trajectories expected to be similar due to a consistent experimental stimulus. DTW's ability to align trajectories despite temporal variations allows for a fair comparison of vehicle responses, supporting robust analysis of following efficiency and energy usage.

\section{Experimental Dataset}\label{sec:dataset}

In this study, the OpenACC dataset is applied to evaluate the effects of EVs and ICEVs equipped with ACC on traffic flow. This section presents an overview of the dataset and describes the data processing steps for analysis.

\subsection{Dataset Overview}

The OpenACC dataset~\citep{makridis2021openacc} provides vehicle trajectory data collected from car-following experiments conducted at five European sites. These experiments are performed on closed test tracks or low-traffic public roads using a standardized protocol to evaluate longitudinal driving behavior across a range of vehicle types and traffic scenarios. The dataset has been widely recognized and used in several studies~\citep{chen2024safety,huang2023characterizing}.

OpenACC includes 10~Hz time-series measurements of vehicle speed, inter-vehicle spacing, and GPS position, along with metadata such as vehicle model, vehicle type (e.g., ICEVs or EVs), driving mode (e.g., human-driven or ACC), ACC distance settings, and road geometry. In each scenario, the lead vehicle operates in ACC mode, while the following vehicle may be either human-driven or under ACC control. This study focuses exclusively on the following vehicles operating in ACC mode in both ICEVs and EVs.

The dataset captures a wide variety of car-following scenarios with up to 27 distinct vehicle models, enabling analysis across diverse driving behaviors. It also reflects a broad spectrum of real-world conditions while preserving experimental control. Equilibrium speeds (i.e., periods where the ACC vehicle maintains a nearly constant speed and gap) range from 6.7--15.6~m/s in urban settings and 13.9--28.2~m/s in highway settings~\citep{yang2024microsimacc}, covering realistic ACC operations across different environments.

\subsection{Data Processing}\label{sec:data-processing}

\begin{figure}[tp!]
    \centering
    \includegraphics[width=1\linewidth]{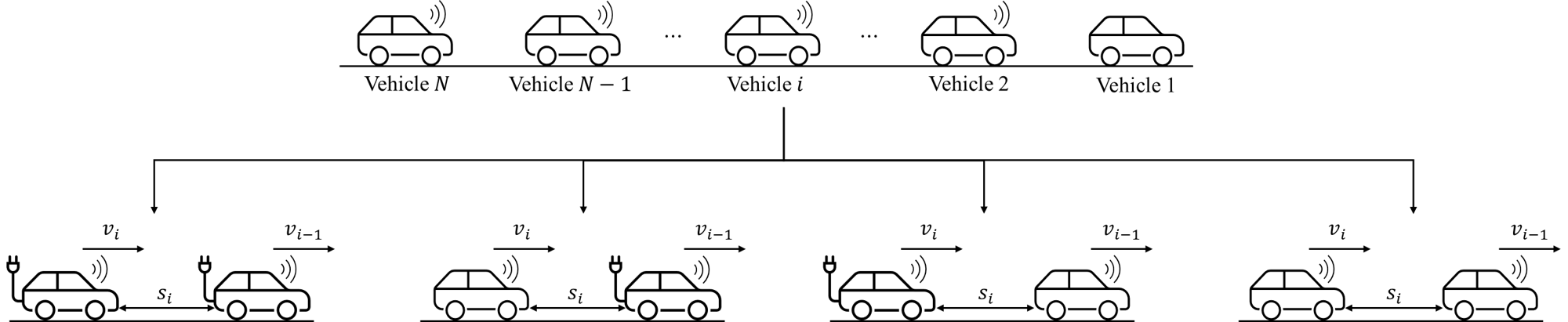}
    \vskip3pt
    \caption{\textnormal{Data processing in the OpenACC dataset. A platoon (top) is decomposed into multiple leader–follower pairs (bottom). Specifically, the bottom row illustrates all four possible pairings, ordered from left to right: EV follower behind EV leader, ICEV follower behind EV leader, EV follower behind ICEV leader, and ICEV follower behind ICEV leader. Each pair serves as a unit of analysis for safety and efficiency evaluations.}}
    \label{fig:platoon-decomp}
\end{figure}

The OpenACC dataset comprises platoons of 2–10 vehicles collected across multiple experimental campaigns~\citep{makridis2021openacc}. In our analysis, we decompose each platoon into consecutive leader and follower pairs and treat each pair as the unit of analysis. For example, an $N$-vehicle platoon (Vehicle 1, 2, $\cdots$, $i$, $\cdots$, $N$) yields $N-1$ pairs. Fig.~\ref{fig:platoon-decomp} illustrates this decomposition process. The top panel shows a complete $N$-vehicle platoon, while the bottom panel shows the resulting leader–follower pairs, clarifying how each follower is associated with the vehicle directly ahead. This visual decomposition highlights that, although platoons move collectively, our analysis focuses on these localized interactions, which provide a consistent basis for comparing efficiency, safety, and environmental effects.

Although the full OpenACC dataset includes a larger number of vehicles and platoons, our analysis applies additional filtering criteria to isolate comparisons between EVs and ICEVs. Specifically, we retain only platoons that include at least one ICEV or EV (where EV refers to battery electric vehicles and excludes hybrids, plug-in hybrids, and fuel cell electric vehicles). Within these filtered platoons, we focus on leader–follower pairs in which the following vehicle is either an EV or an ICEV operating under ACC mode. After applying the filtering criteria, the dataset comprises 59 platoons and 326 leader–follower pairs, including 269 EV followers and 57 ICEV followers. For simplicity, the terms \textit{EV follower} and \textit{ICEV follower} are used to denote the EVs and ICEVs operated by ACC that follow a lead vehicle, respectively. In total, the filtered dataset represents approximately 40 hours of high-frequency trajectory data.

Let the following vehicle be indexed as $i$ and the lead vehicle as $i-1$. The raw trajectory data includes time ($t$), speed ($u$), and IVS (Inter-Vehicle Spacing) of each vehicle with a 10~Hz frequency. Here, IVS is computed from Global Navigation Satellite System (GNSS) data after bumper-to-bumper correction (m). We apply a 10-interval ($w$) moving average to the raw speed $u_i^t$ to suppress high-frequency sensor noise and to stabilize derived features (e.g., acceleration) that are otherwise sensitive to jitter. Concretely, the smoothed speed $v_i^t$ is defined by the moving average:
\begin{eqnarray}
v_i^t = \frac{1}{w}\sum_{z=0}^{w-1} u_i^{t-z}, \quad w = \min\{10,t\},
\end{eqnarray}
where $z$ is the interval index used for the moving average. Here, $z$ ranges from 0 to $w-1$, corresponding to past time intervals from the current time $t$ back to $t-w+1$. At a 10~Hz sampling rate, for every interval $t$, the 10-sample window averages the most recent 1 second of raw speed data. This suppresses noise and spurious spikes while preserving genuine car-following behavior changes.

At each time step $t$ of a leader-follower pair, we extract the following variables for modeling:
\begin{eqnarray}
x_i^t =
\begin{bmatrix}
v_i^t,
v_{i-1}^t,
a_i^t,
a_{i-1}^t,
s_i^t,
\Delta v_i^t
\end{bmatrix},
\end{eqnarray}
where $v_i^t$ and $v_{i-1}^t$ denote the moving average speed of the following and lead vehicle, respectively; $a_i^t$ and $a_{i-1}^t$ are computed using a first-order finite difference method from velocity time series, with a sampling interval of $\Delta t = 0.1$~s:
\begin{eqnarray}
a_i^t = \frac{v_i^{t+\Delta t} - v_i^t}{\Delta t}.
\end{eqnarray}

The spacing $s_i^t$ refers to the bumper-to-bumper headway directly from IVS in the raw data, and the relative speed is given by $\Delta v_i^t = v_{i-1}^t - v_i^t$.

To minimize the influence of startup noise and end-of-run deceleration artifacts, we exclude the beginning and end of each recorded trajectory. Only segments with consistent longitudinal behavior and valid kinematic profiles are retained for further analysis.

\subsection{Trajectory Similarity Analysis}\label{sec:trajectory-similarity-methods}

As discussed in Section~\ref{sec:trajectory-similaity-litreview}, rigorous trajectory matching is essential for conducting reliable empirical analyses. Accordingly, we employ DTW to mitigate potential discrepancies in comparing following-vehicle behaviors that may arise from inconsistencies in the lead-vehicle trajectories.

To conduct DTW, a cost matrix is constructed by computing the Euclidean distance between each point in trajectories $X$ and $Y$. DTW then identifies the optimal alignment between the two sequences by minimizing the cumulative distance over a warping path~\citep{senin2008dynamic}. The recursive DTW distance $D(h, k)$ between two time series, $X = \{x_1, x_2, \cdots, x_m\}$ and $Y = \{y_1, y_2, \cdots, y_n\}$, is based on a recurrence relation. Let $D(h, k)$ represent the accumulated distance of aligning the prefix of $X$ up to index $h$ with the prefix of $Y$ up to index $k$, where $h \in \{1, 2, \cdots, m\}$ and $k \in \{1, 2, \cdots, n\}$ denote time indices in the two respective trajectories. The recurrence relation of the accumulated distance is typically defined as:
\begin{eqnarray}
    D(h, k) = \|x_h - y_k\| + \min \left\{ D(h-1, k), D(h, k-1), D(h-1, k-1) \right\},
\end{eqnarray}
where $\|x_h - y_k\|$ is the local distance between the $h$-th element of $X$ and the $k$-th element of $Y$, e.g., Euclidean distance. When $h=k=1$, the alignment ends at $D(0,0)$, which corresponds to the initialization of the DTW cost matrix, i.e., the cost between the first elements of the two sequences. More details of boundary conditions and initialization can be found in~\citep{senin2008dynamic}. The DTW distance between $X$ and $Y$ is then given by the value at the final cell of the cost matrix, i.e., $\text{DTW}(X, Y) = D(m, n)$.

To account for differences in trajectory length, the final DTW value is normalized by the maximum length of the two sequences:
\begin{eqnarray}
    \text{DTW}_{\text{norm}}(X, Y) = \frac{\text{DTW}(X, Y)}{\max \left\{\text{len}(X), \text{len}(Y) \right\}},
\end{eqnarray}
where $\text{len}(X) = m$ and $\text{len}(Y) = n$ denote the number of time steps (i.e., the length) of trajectories $X$ and $Y$, respectively. For both EVs and ICEVs, we compute DTW pairwise normalized distances between every possible pair of vehicle trajectories, resulting in one DTW distance per trajectory pair. The resulting normalized DTW value serves as a similarity score, where lower values indicate greater similarity between the compared trajectories. A value of zero corresponds to identical trajectories, while larger values reflect less similarity. For each trajectory, we compute the median of its pairwise normalized DTW distances to all other trajectories. This yields one median similarity score per trajectory. We then filter the dataset by retaining the 10 vehicles from each following-vehicle type (EV or ICEV) category with the lowest median DTW values, corresponding to the most similar trajectories. This filtering step reduces the influence of outliers and ensures a consistent basis for upstream efficiency comparisons.

\section{A Framework for Empirical Analysis}\label{sec:framework}

\begin{figure}[tp!]
    \centering
    \includegraphics[width=0.8\textwidth, trim=5 5 5 5, clip]{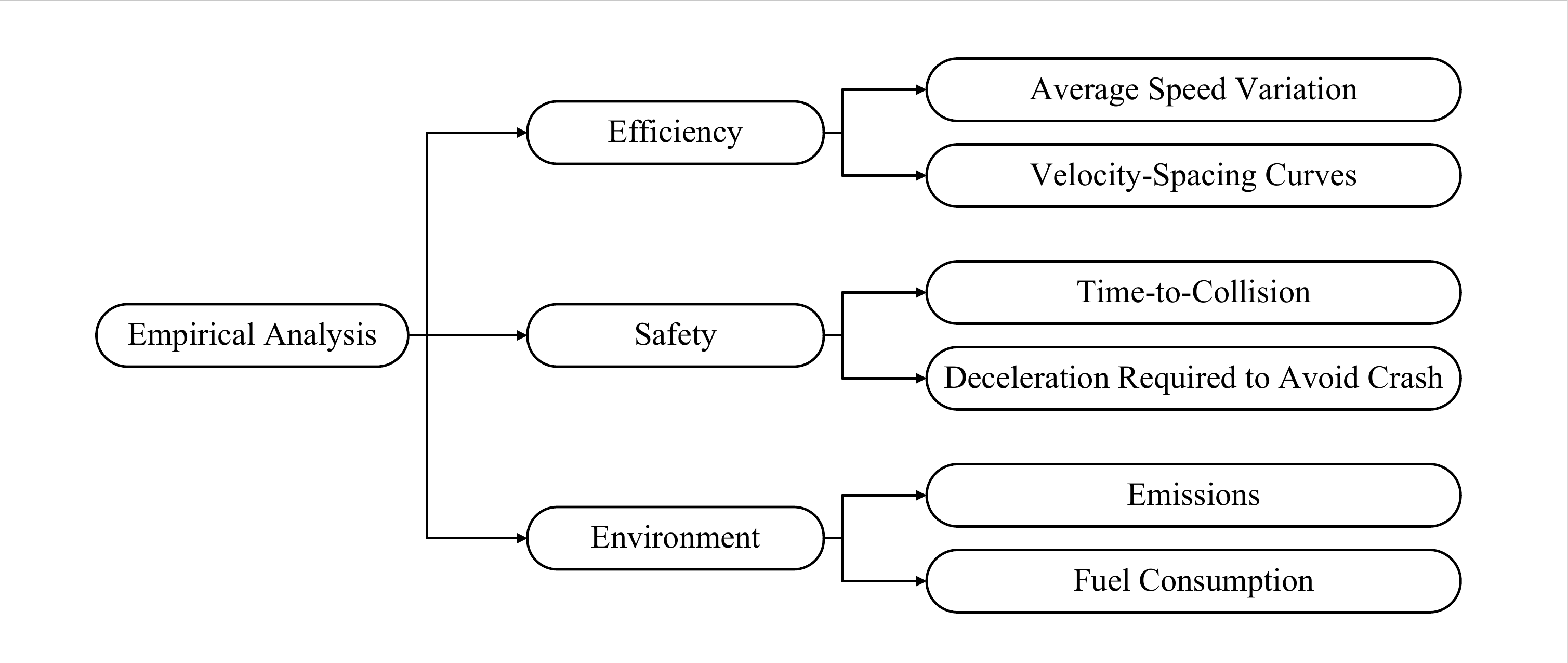}
    \vskip-5pt
    \caption{\textnormal{Analytical framework of the study. The empirical analysis branches into three key impact areas: efficiency, safety, and the environment. Each branch connects to the corresponding quantitative metrics used to evaluate the influence of driver-assist enabled EVs and ICEVs.}}
    \label{fig:study-framework}
\end{figure}

As shown in Fig.~\ref{fig:study-framework}, this study empirically evaluates the impacts of EV adoption across three key dimensions: traffic efficiency, roadway safety, and environmental impacts. Traffic efficiency is assessed through two metrics: average speed variation (ASV) and velocity-spacing ($v$--$s$) curves. Roadway safety is evaluated using time to collision (TTC) and deceleration required to avoid crashes (DRAC), both of which quantify collision risk in vehicle dynamics. Environmental impacts are estimated via the VT-Micro model, which calculates fuel consumption and emissions based on instantaneous speed and acceleration data. 

\subsection{Efficiency Metrics}

To evaluate the impact of EVs and ICEVs on traffic efficiency, we examine two metrics: ASV and $v$--$s$ curves.

\subsubsection{Average Speed Variation (ASV)}

ASV is a metric proposed in~\citep{wang2023general} to quantify traffic oscillations and instabilities. Higher ASV values generally indicate greater fluctuations in speed, which are often associated with stop-and-go behavior, reduced flow stability, and lower traffic mobility. For a given vehicle $i$ observed over the time horizon $[t_1, t_2]$, ASV is defined as:
\begin{eqnarray}
    \text{ASV}_i = \frac{1}{t_{2}-t_{1}} \int_{t_{1}}^{t_{2}} \left| v_i^t - v^* \right| dt,
   \label{eq_ASV}
\end{eqnarray}
where $v_i^t$ is the instantaneous speed of vehicle $i$ at time $t$, and $v^*$ is the vehicle's desired cruising speed. This metric captures deviations from smooth, stable driving behavior.

\subsubsection{Velocity-Spacing ($v$--$s$) Curves}\label{sec:v-s-curves-methods}

To evaluate car-following behavior, we analyze the relationship between a vehicle's instantaneous speed and its spacing to the lead vehicle, commonly referred to as the $v$--$s$ curve~\citep{treiber2013traffic}. The $v$--$s$ relationship forms the basis of many car-following models, such as the Optimal Velocity Model (OVM)~\citep{bando1995dynamical} and the Intelligent Driver Model (IDM)~\citep{treiber2000congested}, and can take either a linear or nonlinear form. For ACC systems, prior studies have shown that this relationship can be linearized, as demonstrated in the Optimal Velocity Relative Velocity (OVRV) model~\citep{milanes2014modeling}. However, the $v$--$s$ curve in the OVRV model does not impose an upper limit on vehicle speed once the desired speed is reached, leading to unrealistic driving behavior. Therefore, we adopt a piecewise linear function that preserves the key characteristics of the OVRV $v$--$s$ relationship while ensuring realism by constraining the vehicle speed not to exceed its desired value.

With this in mind, for each vehicle trajectory, we fit a continuous piecewise linear function to the $v$--$s$ data using the Python library \texttt{pwlf}~\citep{jekel2019pwlf}. The fitting procedure requires the specification of two boundary conditions: jam spacing, which represents the minimum spacing where vehicle speed approaches zero, and desired speed, the asymptotic speed as spacing increases.

\begin{figure}[tp!]
    \centering
    \includegraphics[width=0.5\linewidth]{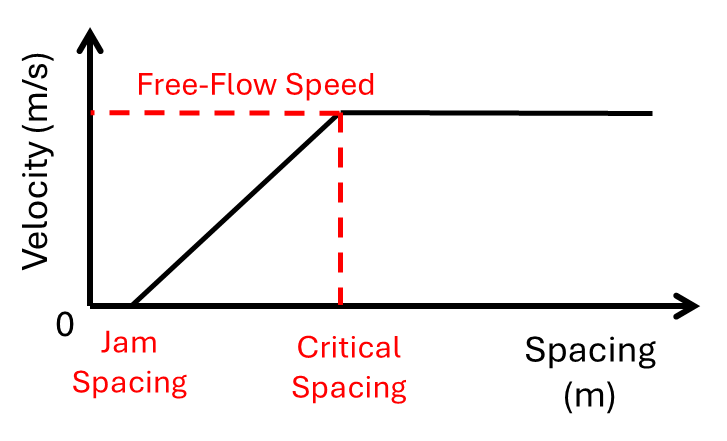} 
    \vskip-5pt
    \caption{\textnormal{Representative schematic velocity–spacing ($v$--$s$) curve. The piecewise linear fit is shown in black. Critical points are highlighted in red: the jam spacing indicates the minimum spacing at which vehicle speed approaches zero, the free-flow speed marks the asymptotic speed as spacing increases, and the critical spacing denotes the transition between car-following and free-flow regimes. This schematic illustrates the general features of $v$--$s$ relationships and car-following behavior.}}
    \label{fig:v-s-curve}
\end{figure}

Fig.~\ref{fig:v-s-curve} illustrates a representative, schematic $v$--$s$ curve. The piecewise linear function, shown in black, captures the relationship of car-following behavior. Key points, including jam spacing, free-flow speed, and critical spacing, are highlighted in red to indicate their roles in defining safe and realistic car-following behavior.

A key parameter derived from the fitted function is the critical spacing, defined as the spacing at which a vehicle transitions from the car-following regime to free-flow conditions. This value reflects the driver's or ACC system's desired time gap and comfort margin in maintaining safe spacing. A smaller critical spacing indicates more aggressive or closely coupled following behavior, whereas a larger value represents more conservative spacing preferences.

\subsection{Safety Metrics}\label{sec:safety-metrics-methods}

To evaluate the safety implications of EV behavior, we use two surrogate safety metrics: TTC and DRAC. These metrics quantify rear-end collision risk under different assumptions: TTC is based on kinematic time horizons, while DRAC considers physical braking requirements.

\subsubsection{Time to Collision (TTC)}\label{sec:ttc-methods}

TTC, a commonly used time-varying metric~\citep{hu2023autonomous,li2020analysis}, quantifies the time remaining before a rear-end collision would occur. At each time step, the TTC between vehicle $i$ and its lead vehicle $i-1$ is calculated as:
\begin{eqnarray}\label{eq_TTC}
    \text{TTC}_{i}^t = \begin{cases}
       \frac{s_i^t}{-\Delta v_i^t},  & ~\text{if}~ \Delta v_i^t < 0, \\
       \infty,  & ~\text{otherwise}.
    \end{cases}
\end{eqnarray}
where $s_i^t$ is the spacing between the two vehicles and $\Delta v_i^t$ is the relative speed, defined as $\Delta v_i^t = v_{i-1}^t - v_i^t$. A smaller TTC value implies a shorter reaction time before a potential collision.

\subsubsection{Deceleration Required to Avoid Crash (DRAC)}

DRAC measures the minimum constant deceleration needed by the following vehicle to avoid a collision, assuming the lead vehicle maintains its current speed~\citep{hu2023autonomous,fu2021comparison}. It is calculated at each time step as:
\begin{eqnarray}\label{eq_DRAC}
    \text{DRAC}_{i}^t = \begin{cases}
       \frac{\left(\Delta v_i^t\right)^2}{s_i^t},  & ~\text{if}~ \Delta v_i^t < 0, \\
       0,  & ~\text{otherwise}.
    \end{cases}
\end{eqnarray}

Higher DRAC values indicate more aggressive braking is required to prevent a collision, implying increased crash risk under realistic deceleration limits.

\subsubsection{Critical Event Rate} \label{sec:crit-rate-calc}

To compare safety outcomes across experiments, we compute the mean proportion of critical events per experiment.  A critical event is defined as any time point where the safety metric exceeds a risk threshold, specifically, when TTC falls below or DRAC rises above a predefined threshold. Following prior literature~\citep{hu2023autonomous}, we use TTC thresholds ranging from 1.0 to 4.0 seconds (in 0.5-second increments) and DRAC thresholds ranging from 2.0 to 5.0 m/s\textsuperscript{2} (in 0.5~m/s\textsuperscript{2} increments).

Formally, the proportion of critical events in a given experiment is calculated as:
\begin{eqnarray}
    p_{\text{crit}} = \frac{c_{\text{crit}}}{c_{\text{total}}} \times 100\%,
\end{eqnarray}
where $c_{\text{crit}}$ is the number of time points where the safety metric indicates risk (e.g., TTC $<$ threshold or DRAC $>$ threshold), and $c_{\text{total}}$ is the total number of evaluated time points. Lower values of $p_{\text{crit}}$ suggest safer driving behavior within the platoon.

\subsection{Environmental Metrics}\label{sec:emissions-methods}

To evaluate the environmental implications of mixed traffic flow, we estimate fuel consumption and tailpipe emissions for ICEV followers within representative platoons. Specifically, we compare platoons led by either an EV or an ICEV under similar traffic conditions. 

Vehicle-level emissions are estimated using the VT-Micro model~\citep{ahn2002estimating}, which relates instantaneous speed and acceleration to fuel consumption and pollutant emission rates via a two-variable exponential regression. The general form of the VT-Micro equation is:
\begin{eqnarray}\label{eq:VT}
\text{MOE}^e = \sum_{r=0}^{3} \sum_{q=0}^{3} C_{rq}^e \cdot v^r \cdot a^q,
\end{eqnarray}
where $\text{MOE}^e$ denotes an instantaneous measure of effectiveness (e.g., fuel consumption or emission rate) corresponding to emission type $e$, $v$ is the vehicle speed in km/h, $a$ is acceleration in km/h/s, and $C_{rq}^e$ are empirically derived model coefficients which are obtained from~\citep{ahn2002estimating}. The coefficients $C_{rq}^e$ are specific to each emission type $e$: fuel consumption, hydrocarbons (HC), carbon monoxide (CO), and nitrogen oxides (NO\textsubscript{x}). For simplicity, the time index $t$ is omitted in Eq.~\eqref{eq:VT}.

To quantify emissions at the platoon level, we compute the average vehicle emission rate across all time steps. Let $\text{MOE}_{i,t}^e$ represent the average instantaneous emission of type $e$ from ICEV follower $i$ at time $t$, where $N$ is the number of ICEV followers in a given platoon and $T$ is the number of time steps. Assuming uniform sampling across vehicles and time steps, the average emission rate, denoted $\text{MOE}^e_{\text{platoon}}$, is given by:
\begin{eqnarray}\label{eq:MOE}
    \text{MOE}^e_{\text{platoon}} = \frac{1}{N \cdot T} \sum_{i=1}^{N} \sum_{t=1}^{T} \text{MOE}_{i,t}^e.
\end{eqnarray}

Eq.~\eqref{eq:MOE} is applied for each emission type $e$. The units of $\text{MOE}^e$ are liters per second for fuel consumption and grams per second for each pollutant, enabling direct comparison of average emission rates between EV-led and ICEV-led platoons.

\section{Numerical Results}\label{sec:results}

This section presents the empirical results corresponding to the three focus areas of the study: efficiency, safety, and emissions. Each subsection summarizes the outcomes associated with specific metrics introduced in Section~\ref{sec:framework}. 

\subsection{Efficiency Results}

\begin{figure}[tp!]
\centering
\subfloat[EV follower.]{
\includegraphics[width=0.4\textwidth]{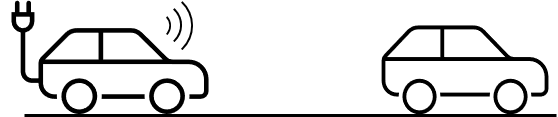}
}
\hfill
\subfloat[ICEV follower.]{
\includegraphics[width=0.4\textwidth]{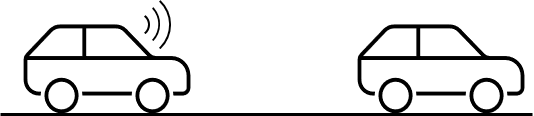}
}
\caption{\textnormal{Leader–follower pairs used for analysis. Each pair includes a lead vehicle and a following vehicle (EV or ICEV) for safety and efficiency evaluations.}}
\label{fig:efficiency-pair-setup}
\end{figure}

This section presents the results of the efficiency analysis, which includes the trajectory similarity assessment and two efficiency metrics: ASV and $v$--$s$ relationships. Recall Section~\ref{sec:data-processing}, the efficiency analysis follows the $v$-$s$ relationship that consists of a leader–follower pair. Fig.~\ref{fig:efficiency-pair-setup} illustrates two scenarios: (a) an ACC-enabled EV following a lead vehicle, and (b) an ACC-enabled ICEV following a lead vehicle.

\subsubsection{Trajectory Similarity Analysis}\label{sec:trajectory-similarity-results}

\begin{figure}[tp!]
  \centering
  \includegraphics[width=0.6\textwidth]{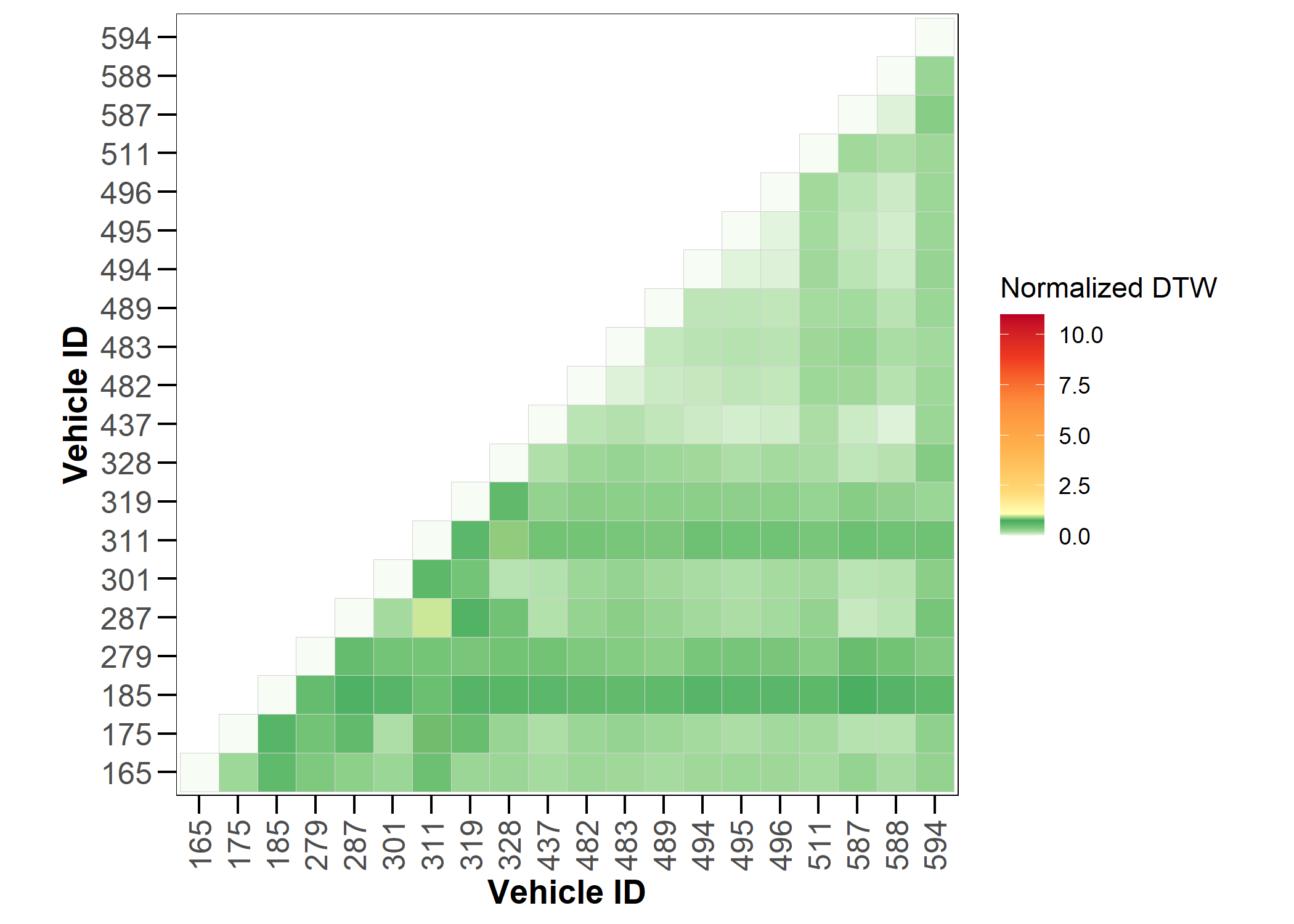}
  \caption{\textnormal{Normalized DTW matrix used to assess trajectory similarity among 20 lead vehicles (10 EVs and 10 ICEVs) for efficiency comparison. Each cell represents the normalized DTW distance between a pair of speed trajectories, with green indicating high similarity (lower distance) and red indicating greater dissimilarity. The matrix is scaled between 0 (identical trajectories) and the maximum DTW value observed among the 277 vehicle pairs evaluated in this study. Due to space constraints, only the subset with the lowest median DTW distances is shown.}}
 \label{fig:dtw-matrix}
\end{figure}

As outlined in Section~\ref{sec:trajectory-similarity-methods}, trajectory similarity analysis is conducted using normalized DTW distances. The resulting distance matrix is shown in Fig.~\ref{fig:dtw-matrix}, where each cell represents the normalized DTW distance between a pair of vehicle speed trajectories.  In the matrix, green indicates higher similarity (i.e., lower DTW distance), while red indicates greater dissimilarity. The matrix is scaled from 0 (identical trajectories) to the maximum DTW value observed across all pairs. To focus on the most representative cases, we display only a subset of results corresponding to the 20 lead vehicles (10 EVs and 10 ICEVs) with the lowest median DTW distances, indicating the highest trajectory similarity among all 277 evaluated vehicle pairs.

\subsubsection{Average Speed Variation (ASV)}\label{sec:asv-results}

\begin{figure}[tp!]
  \centering
  \includegraphics[trim={0cm 0cm 0cm 0cm},clip,width=0.5\textwidth]{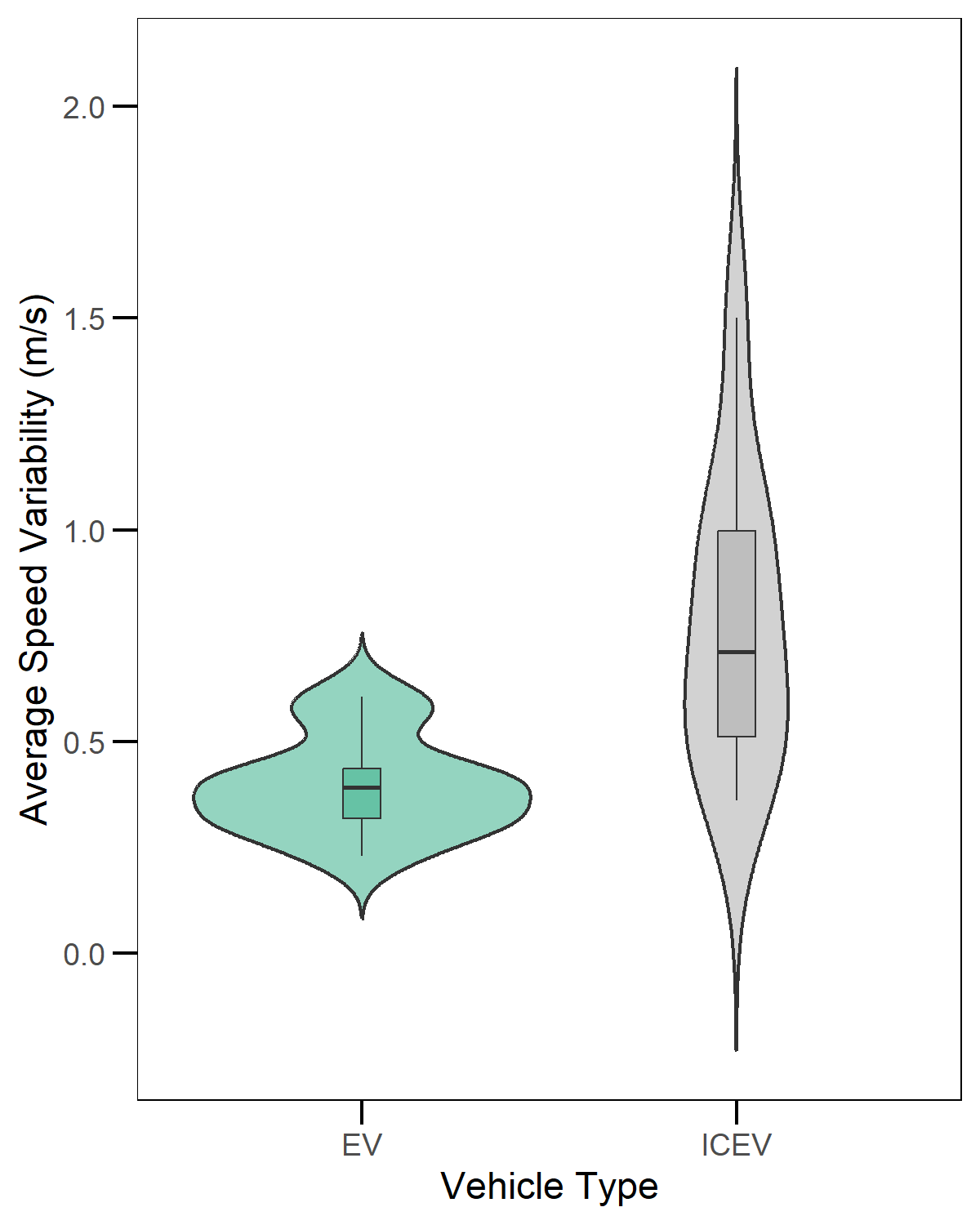}
  \caption{\textnormal{Violin plot of Average Speed Variability (ASV) for EV and ICEV followers. Each distribution is overlaid with a boxplot indicating statistical spread. EV followers (teal) show lower ASV values centered at \(0.393~\text{m/s}\), suggesting smoother driving behavior. ICEV followers (gray) have a broader, skewed distribution with a higher median near \(0.663~\text{m/s}\), indicating greater speed fluctuations.}}
 \label{fig:asv-results}
\end{figure}

ASV is calculated based on the 10 EV followers and 10 ICEV followers filtered from the trajectory similarity analysis results from Section~\ref{sec:trajectory-similarity-results}. These vehicles were identified as having the most representative or typical speed trajectories within their respective platoons, allowing for a meaningful comparison of speed variability across propulsion types. The metric is computed using Eq.~\eqref{eq_ASV}, where a higher ASV value indicates greater speed fluctuations relative to the desired cruising speed.

Fig.~\ref{fig:asv-results} presents a violin plot of ASV values for the two scenarios. The corresponding summary statistics, including minimum, 25th percentile, median, 75th percentile, and maximum values for each group, are reported in Table~\ref{tab:asv_summary}. The distribution for ICEV followers is broader and right-skewed, with a median of 0.663~m/s, suggesting higher speed variability and less stable driving behavior. In contrast, ASV values for EV followers are more tightly clustered around a median of 0.393~m/s, indicating smoother and more stable driving dynamics among ACC-enabled EVs.

\begin{table}[tp!]
\caption{\textnormal{Summary statistics of Average Speed Variability (ASV) by propulsion type, including minimum, 25th percentile, median, 75th percentile, and maximum values. Values correspond to the distribution shown in Fig.~\ref{fig:asv-results}. Lower ASV values indicate much smoother vehicle following.}\label{tab:asv_summary}}
\centering
\renewcommand{\arraystretch}{1.2}
\setlength{\tabcolsep}{8pt}
\begin{tabular}{lcc}
\hline
\noalign{\vskip 1mm}
\textbf{Statistic} & 
\shortstack{\textbf{EV ASV}\\(m/s)} & 
\shortstack{\textbf{ICEV ASV}\\(m/s)} \\
\hline
Minimum & 0.231 & 0.361 \\
25th percentile & 0.317 & 0.503 \\
Median (50th) & 0.393 & 0.663 \\
75th percentile & 0.523 & 1.025 \\
Maximum & 0.607 & 1.500 \\
\hline
\end{tabular}
\end{table}

\subsubsection{Velocity--Spacing ($v$--$s$) Curves}

We apply the modeling approach described in Section~\ref{sec:v-s-curves-methods} to compute and compare $v$--$s$ relationships for ICEV and EV followers. As discussed in Section~\ref{sec:v-s-curves-methods}, the piecewise linear fitting procedure requires specifying boundary conditions, one of which is the jam spacing. We set this boundary condition to 3.57~m, corresponding to the average minimum spacing observed in the dataset. Using the \texttt{pwlf} library~\citep{jekel2019pwlf}, we fit piecewise linear models to the $v$--$s$ data to estimate the critical spacing, which represents the threshold separating the car-following and free-flow regimes.

\begin{figure}[tp!]
  \centering
  \subfloat[ICEV followers. Gray points denote individual spacing–speed data from 10 ICEV follower vehicles. The black line shows the fitted piecewise linear curve.]{
    \includegraphics[width=0.29\textwidth]{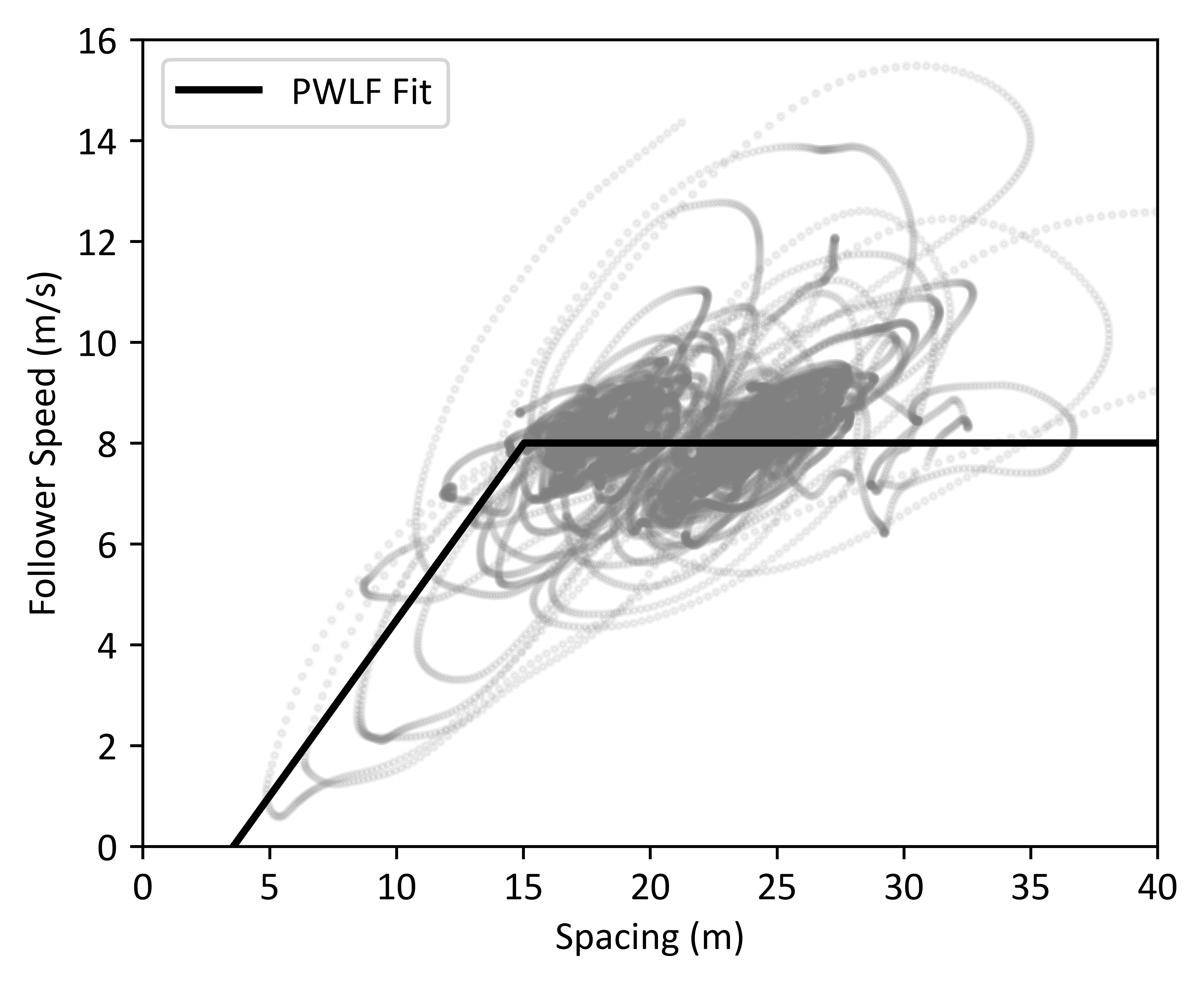}
   \label{fig:icev_vs_curve}
  }
  \hfill
  \subfloat[EV followers. Teal points denote spacing–speed observations from 10 EV followers. The black line represents the fitted piecewise linear $v$--$s$ relationship.]{
    \includegraphics[width=0.29\textwidth]{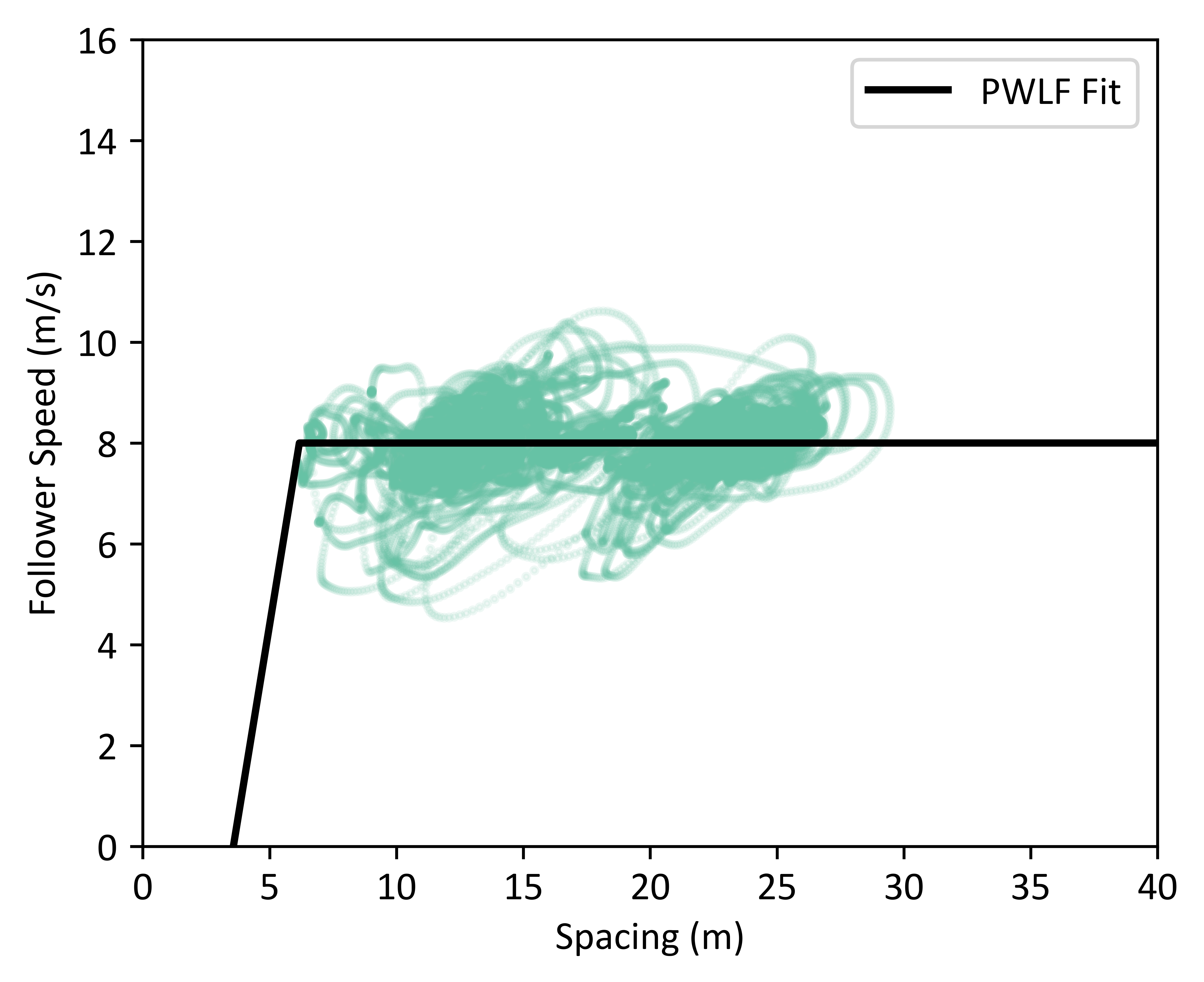}
   \label{fig:ev_vs_curve}
  }
  \hfill
  \subfloat[Comparison of fitted $v$--$s$ curves. The EV curve (solid teal) and ICEV curve (dashed gray) illustrate systematic differences in spacing behavior.]{
    \includegraphics[width=0.29\textwidth]{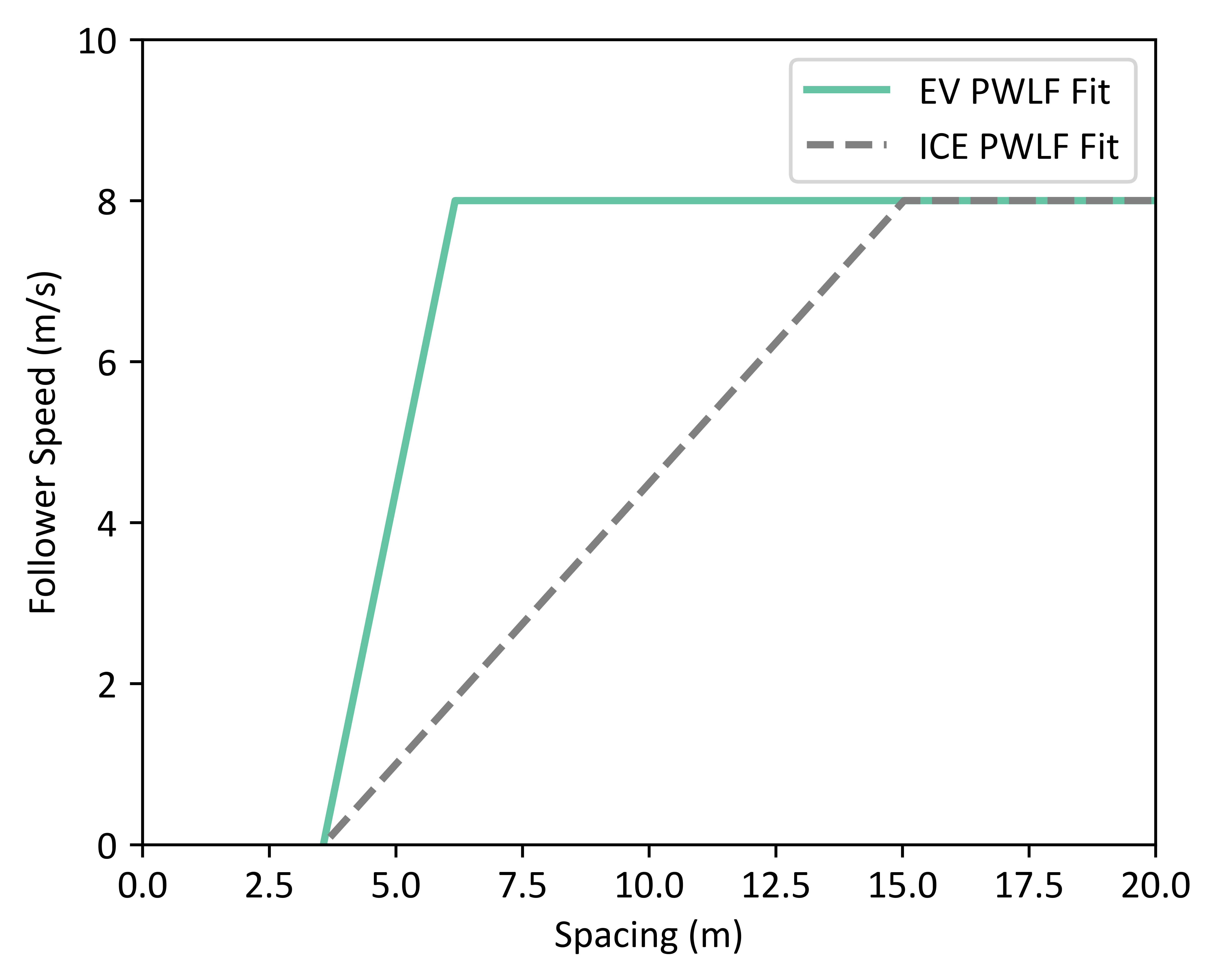}
   \label{fig:comparison_vs_curve}
  }
  \caption{\textnormal{Piecewise linear velocity–spacing ($v$--$s$) relationships for EV and ICEV followers during car-following. Subplots (a) and (b) show the fitted $v$--$s$ curves for each vehicle type, while (c) compares the two fitted curves. Each dot represents an observed spacing–speed pair, and the black lines indicate the fitted piecewise linear functions. The x-axis intersection point represents the jam spacing, the flat segment corresponds to the desired speed, and the transition to the sloped segment identifies the critical spacing.}}
 \label{fig:vs-curves}
\end{figure}

Fig.~\ref{fig:vs-curves} displays the fitted $v$--$s$ curves. Specifically, Fig.~\ref{fig:icev_vs_curve} displays results for ICEV followers, with a critical spacing of 15.03~m. Fig.~\ref{fig:ev_vs_curve} presents the curve for EV followers, with a considerably lower critical spacing of 6.17~m. Fig.~\ref{fig:comparison_vs_curve} provides a direct comparison, showing that EV followers maintain free-flow speeds at shorter headways.

It is noteworthy that ICEV followers exhibit a more conservative car-following pattern, as they tend to decelerate even at relatively larger headways. Although this behavior does not necessarily indicate enhanced safety, it reflects a cautious spacing preference. In contrast, EV followers demonstrate a more compact and space-efficient following behavior, consistent with the ASV results discussed in Section~\ref{sec:asv-results}.

\subsection{Safety Results}

This section presents the results of the safety analysis using two commonly employed surrogate safety metrics: TTC and DRAC. For both metrics, we compute the mean proportion of critical events, which is defined as the proportion of time steps during which the safety metric for a given vehicle falls below a predefined threshold. A higher proportion reflects a greater risk of unsafe or emergency conditions. The safety analysis adopts the same leader–follower pair structure Section~\ref{sec:data-processing}. For reference, the pairwise setup is illustrated in Fig.~\ref{fig:efficiency-pair-setup}.

\subsubsection{Time to Collision (TTC)}

\begin{figure}[tp!]
  \centering
  \includegraphics[width=0.6\textwidth]{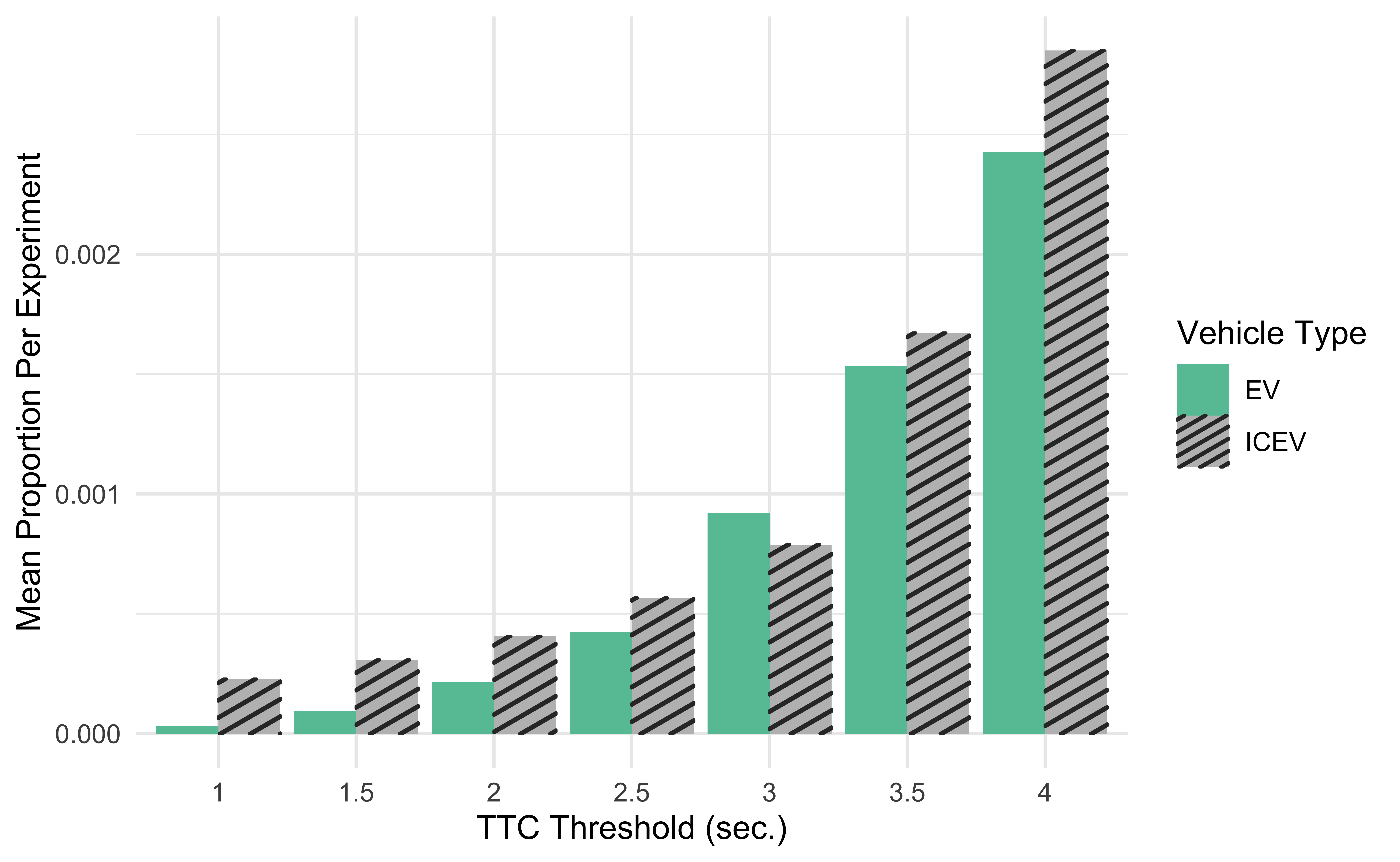}
  \caption{\textnormal{Mean proportion of time steps where the Time-To-Collision (TTC) falls below each threshold, shown along the x-axis. Teal bars represent EV followers and gray bars represent ICEV followers. This analysis captures the relative frequency of potential safety-critical situations for each vehicle type.}}
 \label{fig:ttc-results}
\end{figure}

As described in Section~\ref{sec:crit-rate-calc}, we evaluate the distribution of potentially dangerous following behavior using TTC thresholds ranging from 1.0 to 4.0 seconds, in 0.5-second increments. For each threshold, we compute the mean proportion of time steps at which TTC falls below the given threshold for EV and ICEV followers. The mean proportion of critical events for each TTC threshold are illustrated in Fig.~\ref{fig:ttc-results} and summarized in Table~\ref{tab:ttc_thresholds}.

EV followers exhibit slightly lower proportions of critical events at higher TTC thresholds (3.5--4.0~s), indicating marginally more conservative headway during vehicle following. The largest differences between proportions of critical events for EV and ICEV followers occur at lower TTC thresholds (1.0--2.5~s). For example, at a TTC threshold of 1.0~s, EVs record only 0.0032\% of time steps as dangerous, compared to 0.0228\% for ICEVs, which is an 85.8\% reduction. Similar patterns are observed at the 1.5--2.5~s levels. 

Notably, the results at the TTC threshold of 3.0~s are different from other thresholds, with ICEVs showing a slightly lower proportion of dangerous instances compared to EVs. As this is an empirical study based on real-world driving data, such irregularities may arise due to context-specific factors, sampling variability, or unobserved heterogeneity in the dataset. Given the small magnitude of this deviation and the overall consistency across other thresholds in Table~\ref{tab:ttc_thresholds} and Fig.~\ref{fig:ttc-results}, we interpret this as a potential outlier rather than a reversal of the general trend.

These results suggest that while EVs provide modest improvements at higher TTC thresholds (3.5–4.0~s), they offer substantial safety benefits at lower thresholds (1.0–2.5~s), likely due to the reduced oscillatory flow observed in Section~\ref{sec:asv-results}.

\begin{table}[tp!]
\caption{\textnormal{Mean proportion of TTC critical events below each threshold, disaggregated by propulsion type. Columns show the mean proportion for EV and ICEV followers, while the rightmost column (``Change'') represents the percent difference. Negative values indicate that EV followers experienced a smaller proportion of critical events than ICEV followers, whereas positive values indicate the opposite.}}\label{tab:ttc_thresholds}
\centering
\renewcommand{\arraystretch}{1.2}
\setlength{\tabcolsep}{6pt}
\begin{tabular}{cccc}
\hline
\noalign{\vskip 1mm}
\shortstack{TTC threshold\\(s)} & 
\shortstack{EV\\mean proportion (\%)} & 
\shortstack{ICEV\\mean proportion (\%)} & 
\shortstack{Change \\(\%)} \\
\hline
1.0 & 0.0032 & 0.0228 & $-$85.8 \\
1.5 & 0.0094 & 0.0307 & $-$69.4 \\
2.0 & 0.0217 & 0.0406 & $-$46.6 \\
2.5 & 0.0424 & 0.0566 & $-$25.0 \\
3.0 & 0.0920 & 0.0788 & +16.8 \\
3.5 & 0.1530 & 0.1670 & $-$8.3 \\
4.0 & 0.2430 & 0.2850 & $-$14.8 \\
\hline
\end{tabular}
\end{table}

\subsubsection{Deceleration Required to Avoid Crash (DRAC)}

\begin{figure}[tp!]
  \centering
  \includegraphics[width=0.6\textwidth]{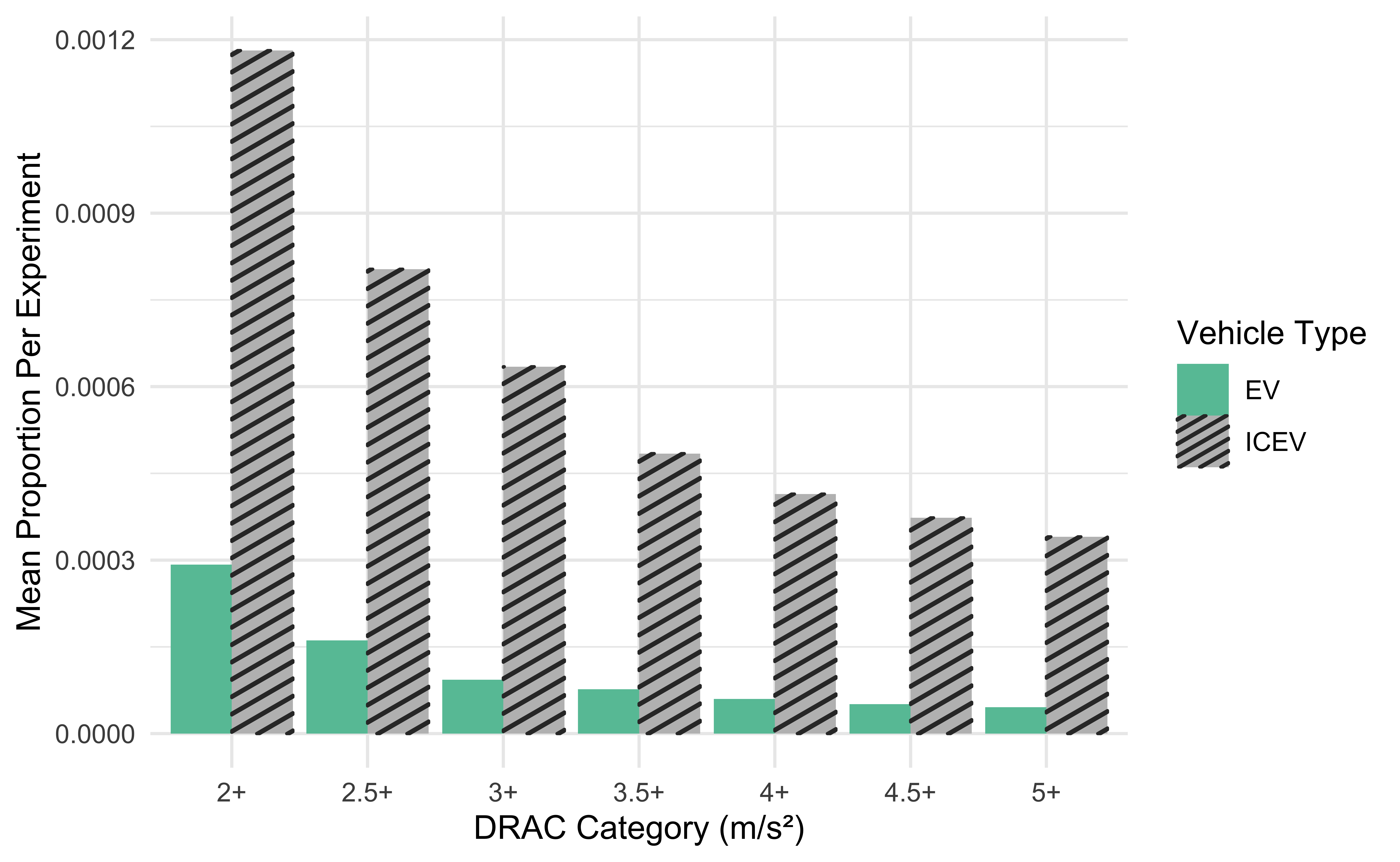}
  \caption{\textnormal{Mean proportion of time steps where the Deceleration Required to Avoid Collision (DRAC) exceeds each threshold, indicating increasing severity of required braking. Teal bars correspond to EV followers, and gray bars correspond to ICEV followers. Higher values suggest more frequent exposure to abrupt braking needs.}}
 \label{fig:drac-results}
\end{figure}

To complement the TTC analysis, we analyze the proportion of time steps where DRAC exceeds thresholds ranging from 2.0 to 5.0~m/s\textsuperscript{2}. As shown in Fig.~\ref{fig:drac-results}, EV followers consistently show lower proportions of high-DRAC instances across all thresholds, a pattern also evident numerically in Table~\ref{tab:drac_thresholds}.

At the most critical threshold of 5.0~m/s\textsuperscript{2}, the dangerous DRAC proportion for ICEVs is 0.0340\%, compared to only 0.0046\% for EVs, which is a reduction exceeding 86.6\%. This performance gap remains significant even at more moderate thresholds, reinforcing the observation that EV followers tend to require less severe braking.

Together, the TTC and DRAC results confirm that EVs improve safety in both normal and emergency driving conditions, defined here as higher TTC thresholds and lower DRAC thresholds for normal conditions, and lower TTC thresholds and higher DRAC thresholds for emergency conditions. These findings support the notion that the smoother profiles of EVs, particularly their regenerative braking and torque delivery, could contribute to more stable and safer platoon behavior.

\begin{table}[tp!]
\caption{\textnormal{Mean proportion of DRAC events above each threshold, disaggregated by propulsion type. Columns show the mean proportion for EV and ICEV followers, while the rightmost column (``Change'') represents the percent difference. Negative values indicate that EV followers experienced a smaller proportion of critical events than ICEV followers, whereas positive values indicate the opposite.}}
\label{tab:drac_thresholds}
\centering
\renewcommand{\arraystretch}{1.2}
\setlength{\tabcolsep}{6pt}
\begin{tabular}{cccc}
\hline
\noalign{\vskip 1mm}
\shortstack{DRAC threshold\\(s)} & 
\shortstack{EV\\mean proportion (\%)} & 
\shortstack{ICEV\\mean proportion (\%)} & 
\shortstack{Change\\(\%)} \\
\hline
2.0+ & 0.0292 & 0.1180 & $-$75.3 \\
2.5+ & 0.0161 & 0.0803 & $-$79.9 \\
3.0+ & 0.0093 & 0.0634 & $-$85.3 \\
3.5+ & 0.0077 & 0.0484 & $-$84.1 \\
4.0+ & 0.0060 & 0.0414 & $-$85.6 \\
4.5+ & 0.0051 & 0.0373 & $-$86.4 \\
5.0+ & 0.0046 & 0.0340 & $-$86.6 \\
\hline
\end{tabular}
\end{table}

\subsection{Environmental Results}

\begin{figure}[tp!]
  \centering
  \subfloat[ICEV-led platoon. Platoons A and B correspond to ICEV-led platoon configurations.\label{fig:icev-led-platoon}]{
    \includegraphics[width=0.45\textwidth]{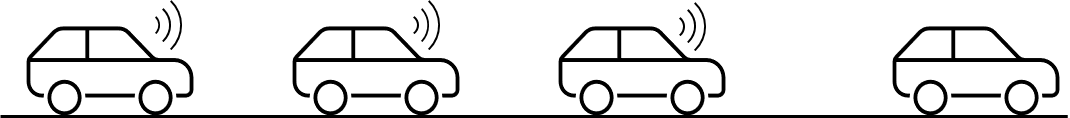}
  }
  \hfill
  \subfloat[EV-led platoon. Platoons C and D correspond to EV-led platoon configurations.\label{fig:ev-led-platoon}]{
    \includegraphics[width=0.45\textwidth]{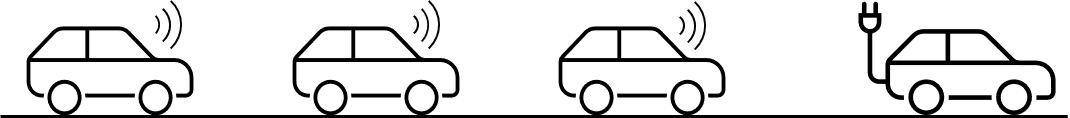}
  }
  \caption{\textnormal{Schematic of the platoon structure used in environmental modeling. Unlike the pair-based setup used for safety and efficiency analysis, the environmental analysis aggregates emissions across all ICEV followers in a platoon, grouped by the propulsion type of the lead vehicle (EV or ICEV).}}
  \label{fig:platoon-emissions-setup}
\end{figure}

As described in Section~\ref{sec:emissions-methods}, the VT-Micro model is employed to assess the environmental impacts of platoon leaders by comparing emissions from ICEV followers led by either an EV or an ICEV. Fig.~\ref{fig:platoon-emissions-setup} illustrates the analysis framework, which evaluates the average emissions of ICEV followers across entire platoons, grouped according to the propulsion type of the lead vehicle. In the results that follow, Platoons A and B refer to separate, identical instances of the same ICEV-led platoon configuration shown in Fig.~\ref{fig:icev-led-platoon}, while Platoons C and D refer to separate, identical instances of the same EV-led platoon configuration shown in Fig.~\ref{fig:ev-led-platoon}. It should be noted that this analysis framework differs from the leader–follower pair configuration adopted in the preceding safety and efficiency analyses.

\begin{figure}[tp!]
  \centering
  \subfloat[Fuel Consumption.]{
    \includegraphics[width=0.4\textwidth]{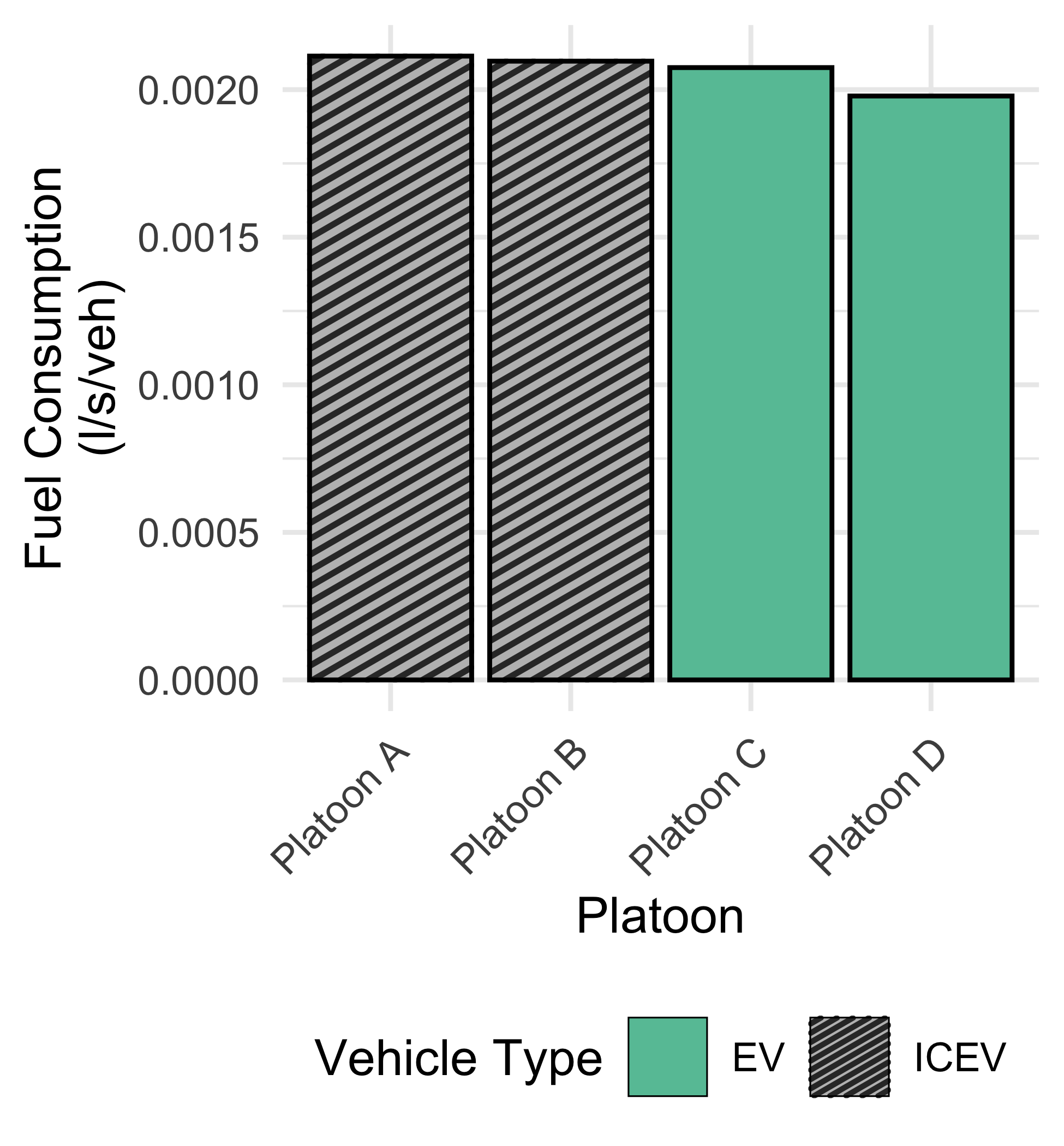}
   \label{fig:fuel}
  }
  \subfloat[CO emissions.]{
    \includegraphics[width=0.4\textwidth]{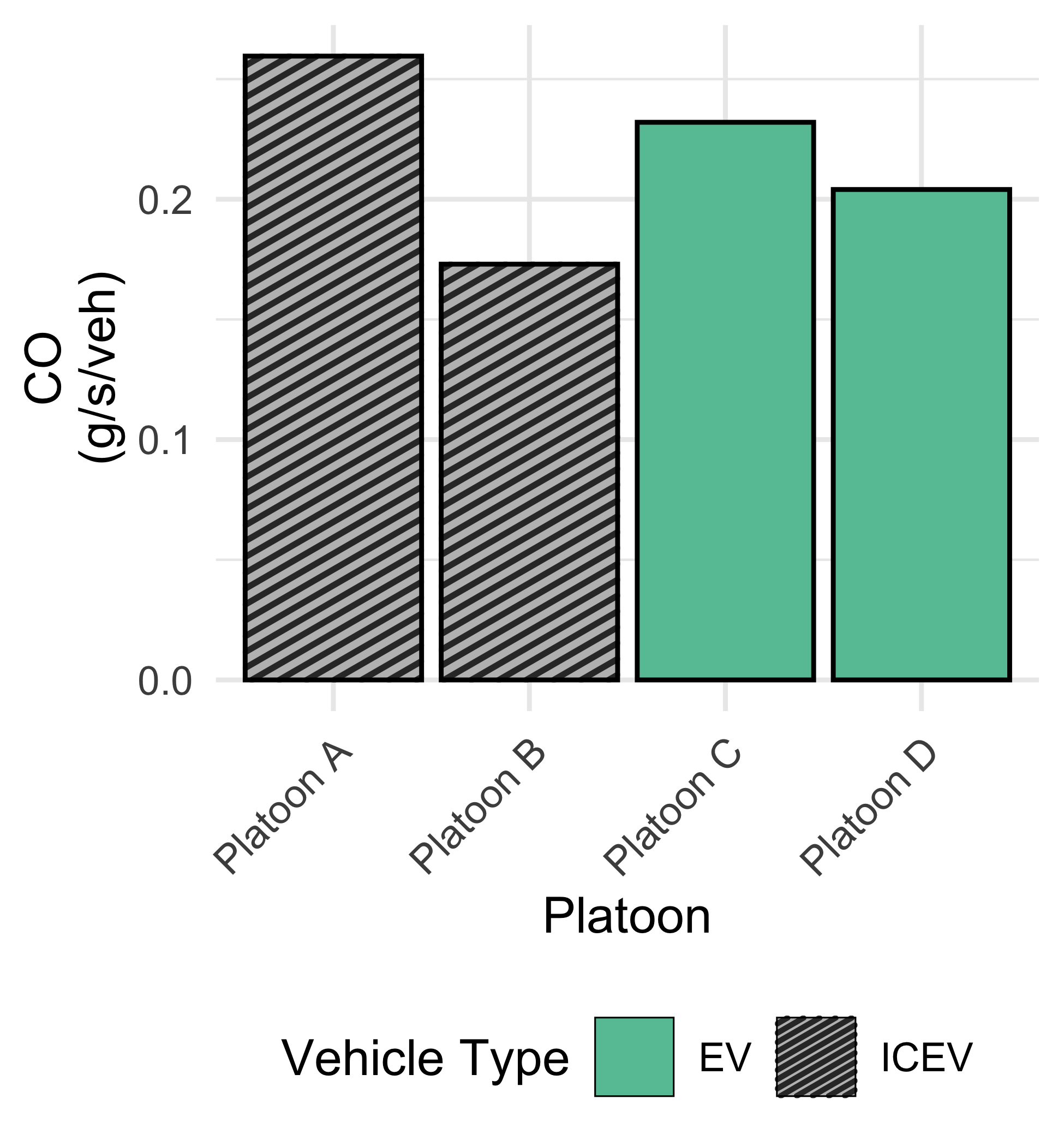}
   \label{fig:co}
  }\\
  \subfloat[HC emissions.]{
    \includegraphics[width=0.4\textwidth]{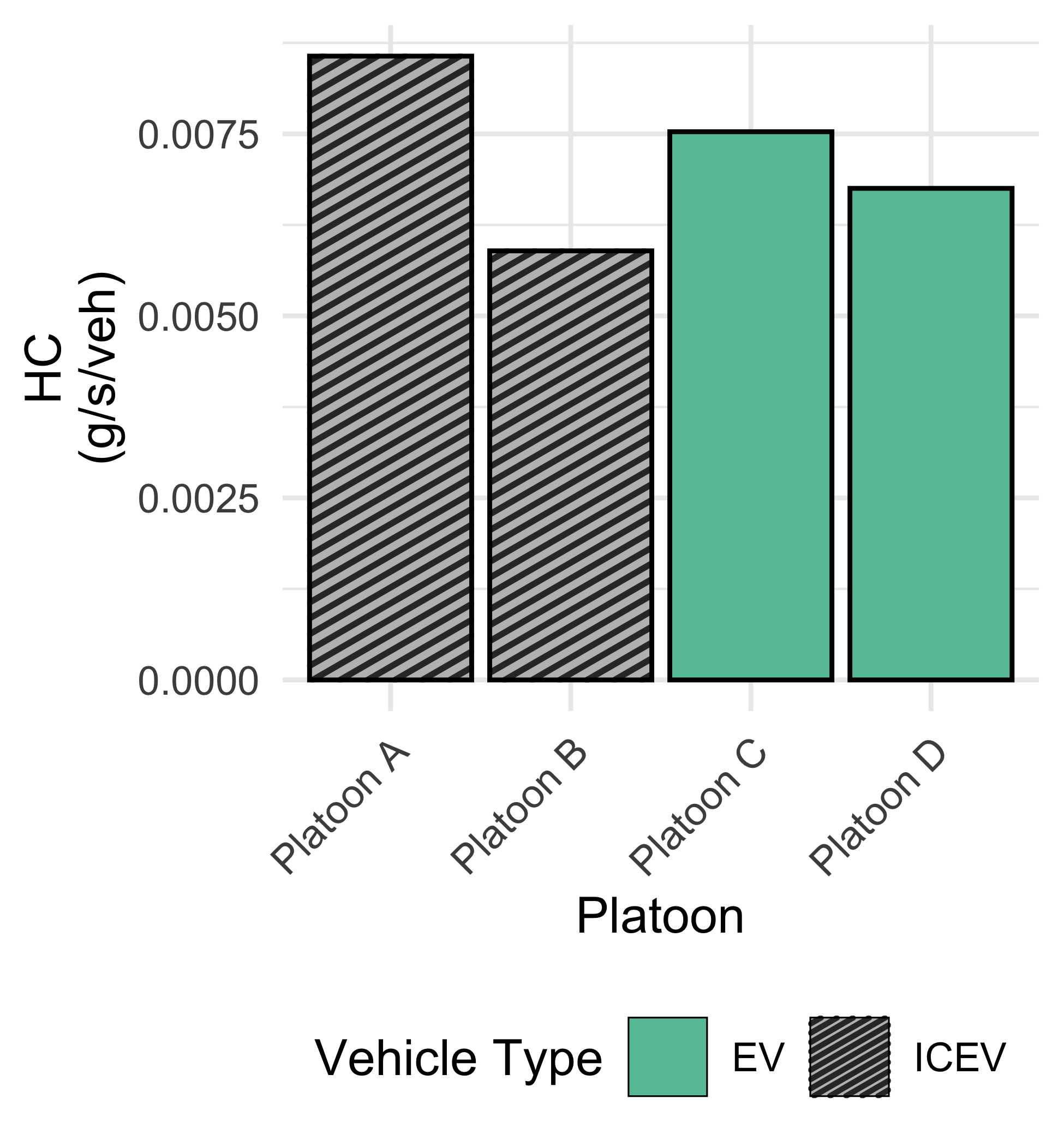}
   \label{fig:hc}
  }
  \subfloat[NO$_\text{x}$ emissions.]{
    \includegraphics[width=0.4\textwidth]{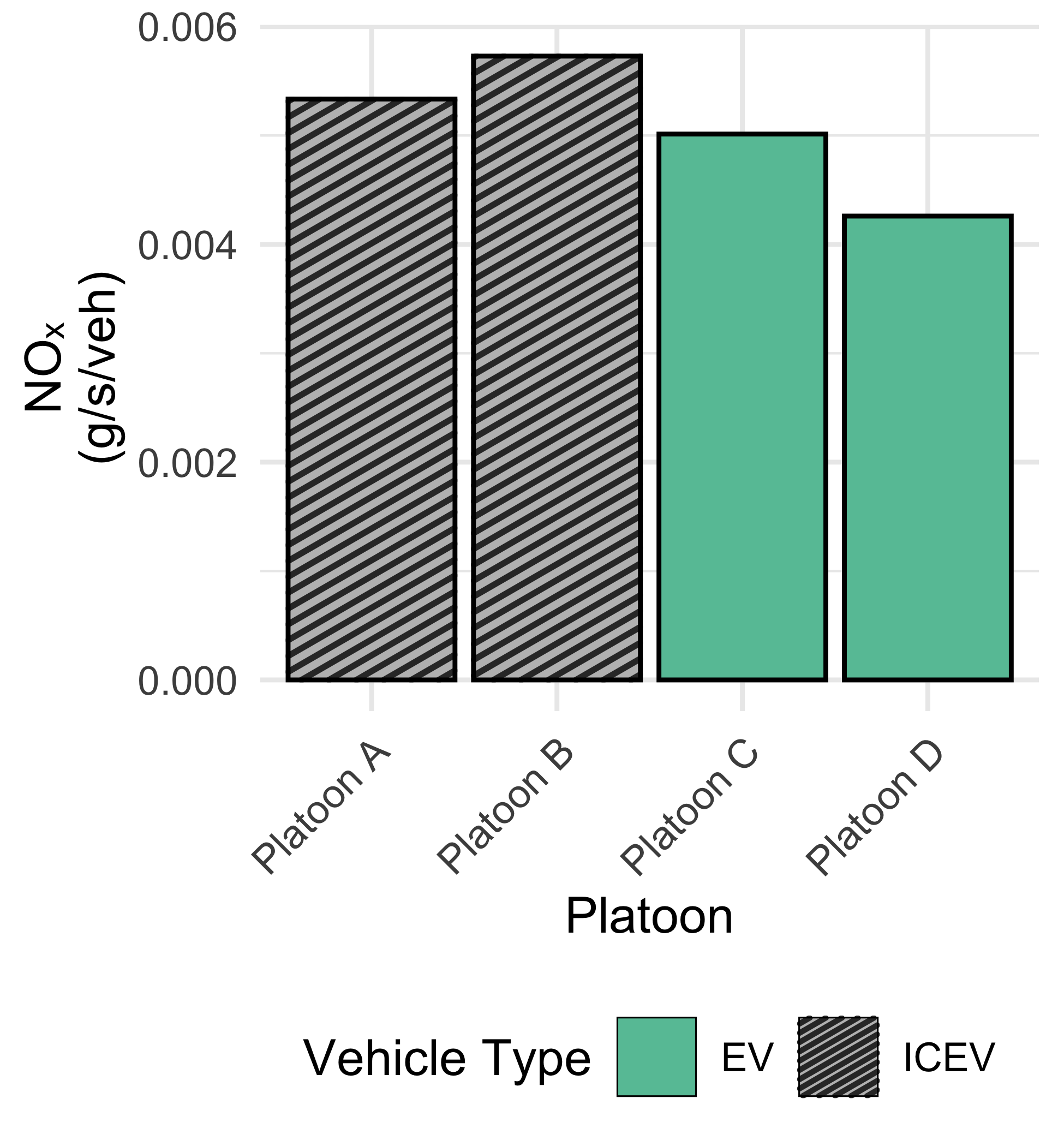}
   \label{fig:nox}
  }
  \caption{\textnormal{Environmental impact comparison of EV-led versus ICEV-led platoons across four emissions-related metrics: fuel consumption, carbon monoxide (CO), hydrocarbon (HC), and nitrogen oxides (NO$_\text{x}$). Platoons A and B correspond to ICEV-led platoons, while Platoons C and D correspond to EV-led platoons. Teal boxplots represent EV-led platoons and gray represent ICEV-led platoons.}}
 \label{fig:emissions}
\end{figure}

\begin{table}[tp!]
    \caption{\textnormal{Average per-vehicle emission rates for each platoon, disaggregated by fuel consumption and emissions type (HC, CO, NO$_\text{x}$). The bottom rows report percent changes in average per-vehicle emission rates for all defined ICEV-led to all EV-led platoon comparisons.}}
    \label{tab:platoon_emissions}
    \centering
    \begin{adjustbox}{max width=\linewidth}
    \begin{tabular}{lccccc}
        \hline
        \noalign{\vskip 1mm}
        Platoon & Lead & \shortstack{Fuel Consumption\\(liters/sec/veh)} 
        & \shortstack{HC\\(grams/sec/veh)} 
        & \shortstack{CO\\(grams/sec/veh)} 
        & \shortstack{NO$_\text{x}$\\(grams/sec/veh)} \\
        \hline
        Platoon A & ICEV & 2.11$\times$10$^{-3}$ & 8.56$\times$10$^{-3}$ & 0.260 & 5.33$\times$10$^{-3}$ \\
        Platoon B & ICEV & 2.10$\times$10$^{-3}$ & 5.89$\times$10$^{-3}$ & 0.173 & 5.73$\times$10$^{-3}$ \\
        Platoon C & EV & 2.07$\times$10$^{-3}$ & 7.53$\times$10$^{-3}$ & 0.232 & 5.01$\times$10$^{-3}$ \\
        Platoon D & EV & 1.98$\times$10$^{-3}$ & 6.75$\times$10$^{-3}$ & 0.204 & 4.23$\times$10$^{-3}$ \\
        \hline
        \shortstack{Change: A $\rightarrow$ C (\%)} & -- 
        & -1.9 & -12.0 & -10.8 & -6.0 \\
        \shortstack{Change: A $\rightarrow$ D (\%)} & -- 
        & -6.2 & -21.1 & -21.5 & -20.6 \\
        \shortstack{Change: B $\rightarrow$ C (\%)} & -- 
        & -1.4 & +27.8 & +34.1 & -12.6 \\
        \shortstack{Change: B $\rightarrow$ D (\%)} & -- 
        & -5.7 & +14.6 & +17.9 & -26.2 \\
        \hline
    \end{tabular}
    \end{adjustbox}
\end{table}

The exact average per-vehicle emission rates for each platoon are summarized in Table~\ref{tab:platoon_emissions}. Results are visualized in Fig.~\ref{fig:emissions}, which shows the corresponding distributions for fuel consumption, HC, CO, and NO\textsubscript{x}. Overall, EV-led platoons consistently demonstrate lower emission rates across most categories and scenarios, as shown in Table~\ref{tab:platoon_emissions}. For fuel consumption, EV-led platoons reduce average consumption by 1.4\% to 6.2\% relative to ICEV-led counterparts. Reductions are more substantial for HC emissions, with average rates declining by 12.0\% to 21.1\%. Although one ICEV-led platoon (Platoon B) exhibits slightly lower HC emissions than the EV-led cases, the broader trend still favors EV-led scenarios. A similar pattern is observed for CO emissions. While Platoon B again presents a minor exception, EV-led platoons generally achieve reductions of 10.8\% to 21.5\%. For NO\textsubscript{x}, the trend is unambiguous: EV-led platoons outperform ICEV-led platoons in all experiments, with emission reductions ranging from 6.0\% to 26.2\%.

These findings suggest that EVs not only offer safety and efficiency benefits as platoon leaders, but also produce indirect environmental benefits by moderating the emissions behavior of trailing ICEVs. This effect is likely attributable to the smoother driving patterns induced by EVs. 

\section{Conclusions}\label{sec:conclusions}

The increasing integration of automation and electrification in modern vehicles presents both opportunities and challenges for the future of transportation. While ACC-equipped vehicles are widely viewed as precursors to full automation, growing evidence suggests that their influence on traffic flow, safety, and energy efficiency may not always be beneficial, especially when operating with traditional internal combustion engine powertrains. At the same time, EVs introduce distinct performance characteristics, such as regenerative braking and high torque availability, which may complement or even mitigate the limitations of existing driver assistance systems. This study addressed an important gap in the literature by empirically evaluating the interaction between vehicle powertrain and ACC behavior under real-world driving conditions. Leveraging the large-scale OpenACC dataset, we conducted a comparative analysis of ACC-enabled EVs and ICEVs, focusing on safety, efficiency, and environmental performance. The use of DTW-based trajectory similarity to align follower scenarios in uncontrolled field settings, enables rigorous and fair comparisons across vehicle types. 

Our findings reveal consistent performance advantages for EVs across safety, efficiency, and environmental metrics. EV followers exhibited significantly lower ASV and a tighter distribution of values, indicating smoother and more stable driving behavior. $v$--$s$ curves showed that EVs maintained free-flow speeds at shorter headways, with a critical spacing nearly 60\% smaller than that of ICEVs, suggesting more agile and responsive following. Safety outcomes further reinforce these trends: EVs demonstrated an 86\% reduction in high-risk TTC and DRAC instances compared to ICEVs at the most critical thresholds. Finally, environmental modeling using VT-Micro indicated that EV-led platoons reduced ICEV follower emissions across fuel consumption, HC, CO, and NO\textsubscript{x}, by up to 26.2\%, highlighting indirect sustainability benefits in mixed platoons. Together, these findings suggest that electrification can partially offset the performance challenges previously associated with ACC systems, leading to more stable, safer, and cleaner traffic flow. The results emphasize the importance of evaluating automation technologies not in isolation, but in conjunction with powertrain design and vehicle energy systems.

While this study provides key insights, it is based on a limited dataset of EV driving behavior, and the findings should be interpreted within this context. Nevertheless, the proposed methodology integrates trajectory similarity analysis and a structured evaluation framework that is generalizable and can be applied to larger and more diverse datasets as they become available. As EV penetration increases and richer naturalistic datasets are collected, the framework developed here can enable deeper and more comprehensive assessments of the interplay between automation, electrification, and traffic performance.

This study focuses primarily on longitudinal vehicle dynamics, abstracting away lateral behaviors such as lane changing to maintain analytical clarity. However, in real-world traffic, maneuvers like lane changing, merging, and cut-ins introduce perturbations that can significantly affect car-following behavior and traffic stability. Future research should explicitly incorporate these lateral dynamics and examine their interaction with longitudinal control, particularly in mixed traffic environments where human-driven and automated vehicles coexist.

\bibliographystyle{unsrtnat}
\bibliography{reference}

\end{document}